\documentclass[12pt]{article} 

\usepackage{amsmath,amssymb}
\usepackage[dvips]{graphicx,color}
\usepackage[round]{natbib}
\usepackage{mathrsfs}
\usepackage{bm, xcolor}
\usepackage{mathrsfs}
\usepackage{verbatim}
\usepackage{threeparttable}
\usepackage{color}
\usepackage[colorlinks,
linkcolor=blue,
anchorcolor=blue,
citecolor=blue]{hyperref}
\usepackage[utf8]{inputenc}
\usepackage{epstopdf}
\usepackage{booktabs}
\usepackage{graphicx}
\usepackage{subfigure}
\usepackage{float}
\usepackage{listings}
\usepackage{amssymb}
\usepackage{amsthm}
\usepackage{multirow}
\usepackage{rotating}
\usepackage{longtable}
\usepackage{enumitem}
\usepackage{threeparttablex} 
\usepackage{booktabs} 

\usepackage{cases}

\usepackage{caption}
\usepackage{filecontents}
\usepackage{algorithmic}
\usepackage{algorithm}
\usepackage{float}

\allowdisplaybreaks
\def\P{\mathbb{P}}
\def\E{\mathbb{E}}

\def\o{{\scriptstyle{\mathcal{O}}}}
\def\O{\mathcal{O}}
\def\i{\mathrm{i}}
\def\op{\mathrm{op}}

\newtheorem{theorem}{Theorem}[section]

\newtheorem{remark}{Remark}[section]

\numberwithin{equation}{section}
\usepackage{chngcntr}

\newcommand{\blind}{0}

\begin{document}

\def\spacingset#1{\renewcommand{\baselinestretch}%
{#1}\small\normalsize} \spacingset{1}

\date{}
\if0\blind
{
  \title{\bf Distributional change point analysis: high dimensionality, nonstationary dependence, and computational constraints}
  \author{Leheng Cai\footnotemark[1] \footnotemark[4],  \,  Moshu Xu\footnotemark[1] \footnotemark[4] \, and  \, Zhou Zhou\footnotemark[2] \footnotemark[3] \footnotemark[4]\hspace{.2cm}}
  \maketitle
  \renewcommand{\thefootnote}{\fnsymbol{footnote}}

  \footnotetext[1]{Department of Statistics and Data Science, Tsinghua University}
           
		\footnotetext[2] {Center for Data Science and School of Management, Zhejiang University}
        \footnotetext[3] {Corresponding author: zhouzhou.stat@gmail.com}
        \footnotetext[4] {All authors contributed equally  to this article and  are listed in the alphabetical order.}
} 

\if1\blind
{
  \bigskip
  \bigskip
  \bigskip
  \begin{center}
    { \bf }
\end{center}
  \medskip
} \fi

\bigskip

\maketitle
	
	\textbf{Abstract:}   
We study general distributional change point analysis for high-dimensional
time series under nonstationary temporal dependence and limited computational
resources. To mitigate noise 
accumulation from irrelevant coordinates and reduce the computational cost, we propose a sparse random projection procedure based on characteristic functions. We   quantify the computational
complexity of the random projections and show that the required number
of projections grows polynomially with the ambient dimension. 
To accommodate 
nonstationary temporal dependence, a Gaussian multiplier bootstrap
based on contrasts of local empirical characteristic functions is developed. We
establish asymptotic false detection control and simultaneous guarantees for
estimating the number and locations of multiple change points, with a near-parametric  uniform
localization rate up to
logarithmic factors. 
A key technical innovation is a Gaussian approximation for high-dimensional
multi-scale sparse time series, which is of independent interest. Simulation studies demonstrate the effectiveness of the proposed procedure in  detecting and localizing distributional changes under various temporal dependence structures. An application to neural recordings further illustrates the practical utility of the proposed method in detecting diverse distributional changes in neural activity.


	\maketitle
	
\newpage
\spacingset{1.5} 

\section{Introduction}

Change point analysis provides a fundamental framework for identifying structural
instabilities in data observed sequentially over time and has a long history in
statistics, econometrics, and signal processing; see, among others,
\citet{basseville1993detection},  and
\citet{csorgo1997limit} for comprehensive
treatments. With the increasing availability of large-scale multivariate time
series, change point problems in which the dimension is comparable to, or even larger than, the sample size have received considerable attention;
see \citet{liu2022high} for a recent review.

An important challenge in such applications is that the form of a structural
change is often unknown in advance. A change may occur in the mean, covariance,
tail behavior, skewness,  or other features of the distribution,
while some low-order characteristics remain unchanged; see
the neural recordings analyzed in Section~\ref{sec:realdata} for illustration. The detected changes include jumps in individual units' mean firing rates, changes in pairwise covariance and joint upper-tail behavior, and shifts in active-bin proportions and mean counts per active bin,  motivating distributional change point analysis that accommodates different forms of change without specifying them in advance.

Consider a  $p$-dimensional time series
$\{\bm X_i\}_{i=1}^{n}$, where the dimension is allowed to depend on the sample
size and diverge, that is, $p=p_n\to\infty$ as $n\to\infty$. Let $F_i$ denote
the marginal distribution of $\bm X_i$. For
\(
0=\tau_0<\tau_1<\cdots<\tau_{r_n}<\tau_{r_n+1}=n,
\)
suppose that
\(
F_{\tau_k+1}=\cdots=F_{\tau_{k+1}}\),
for \(k=0,1,\ldots,r_n\),
so that the one-time marginal distribution remains constant within each segment
$(\tau_k,\tau_{k+1}]$ and changes at
$\tau_1,\ldots,\tau_{r_n}$. The number  of change points $r_n$ is allowed to diverge with sample size $n$. Define the minimum spacing between consecutive
change points by
\(
\vartheta_n
:=
\min_{0\le k\le r_n}
(\tau_{k+1}-\tau_k).
\)
Our first goal is to test $H_0:r_n=0$ versus $H_1:r_n\ge1$. 
If the null hypothesis is rejected, we further aim to estimate both the number
$r_n$ of change points and their locations
$\tau_1,\ldots,\tau_{r_n}$. 

A substantial literature has developed nonparametric methods for detecting distributional 
changes. Representative approaches are based on
empirical characteristic functions \citep{huskova2006change}, empirical
distribution functions \citep{zou2014nonparametric}, energy distances
\citep{matteson2014nonparametric},  kernel embeddings
\citep{arlot2019kernel}, and nonparametric density estimation \citep{madridpadilla2023change}.  These methods avoid specifying in advance which
particular feature of the distribution changes and are therefore attractive
against heterogeneous alternatives. However, their theoretical developments are primarily concerned with fixed-dimensional settings
and do not address the statistical difficulties when the dimension $p$ diverges.

Much of the literature on high-dimensional change point analysis targets
prespecified distributional features, most notably means, variances, or
covariances; see, for example, \citet{jirak2015uniform}, \citet{wang2018high},  
\citet{enikeeva2019high}, and \citet{liu2020unified}. 
Such procedures can be highly effective when the form of the change is known,
but they are typically sensitive only to a limited class of distributional
changes associated with the targeted features, and have little power for other types of changes.
Recently, several methods have addressed general distributional changes in high dimensions. \citet{chen2015graph} introduced graph-based scan
statistics for multivariate and non-Euclidean observations, and
\citet{chu2019asymptotic} developed related asymptotically distribution-free
procedures with improved sensitivity to a broader range of alternatives.
\citet{song2022practical} further developed a more efficient procedure using kernel-based discrepancies.
\citet{chakraborty2026high} proposed a procedure based on generalized
homogeneity metrics to detect high-dimensional distributional changes beyond the
first two moments. \citet{boniece2024change} developed $U$-statistic-based
procedures for distributional changes in high dimensions, while
\citet{drikvandi2025distribution} proposed a distribution-free distance-based
method particularly suited to non-sparse high-dimensional alternatives.
These developments broaden the scope of high-dimensional
change point analysis beyond mean and covariance changes. However, retaining computational efficiency in high dimensions still remains challenging for most existing procedures. 

Another common limitation of the existing literature on high-dimensional distributional change point analysis is its reliance on the assumption of mutually independent observations. This assumption can be overly restrictive for temporally ordered data where serial dependence is often intrinsic. Furthermore, non-stationarity is the rule rather than the exception in modern complex temporal systems. Even when the marginal distribution remains unchanged, the underlying temporal dependence structure may evolve over time, rendering stationarity assumptions unnecessarily restrictive \citep{dahlhaus1997fitting}. For instance, temporal correlations in neural activity have been shown to vary with attention \citep{zeraati2023intrinsic}. Motivated by these considerations, our methodology is developed under a general nonstationary framework that allows the temporal dynamics to evolve independently of changes in the marginal distribution. Specifically, while the marginal distribution of $\bm X_i$ remains constant between consecutive change points, the underlying data-generating mechanism and serial dependence  are permitted to vary over time. Consequently, even under the null $H_0$ of no distributional change, the time series need not be stationary. A formal characterization of the admissible nonstationary temporal dependence is provided in Section 3. 
This formulation differs from approaches that treat changes in serial
dependence as signals to be detected (e.g., \citet{dette2019change};
\citet{mcgonigle2025nonparametric}).


The high-dimensional distributional change point problem under nonstationarity gives rise to several challenges. 
First, general distributional changes lack a finite-dimensional parameterization, calling for an omnibus representation that remains informative in high dimensions. However, classical omnibus discrepancies can become dominated by low-order
distributional features as the dimension increases
\citep{ramdas2015decreasing,yan2023kernel}. 
Second, informative distributional signals may be obscured by many irrelevant
coordinates, while direct comparison of $p$-dimensional distributions becomes
increasingly challenging as $p$ grows. {Third, the computational difficulty is also intrinsic in high dimensions. Even after restricting
attention to sparse informative directions,
naive optimization over such directions is
NP-hard in general; see \cite{wang2018high}.}
Finally,
nonstationary temporal dependence significantly complicates calibration.
Even when the marginal distribution is constant under $H_0$, the covariance structure of
the resulting change point statistics may vary with time, so that calibration
methods based on independence or  piecewise stationarity are no longer directly applicable.

We address the first three difficulties by using characteristic functions and sparse random projections. Two
probability distributions are identical if and only if their characteristic
functions agree at all frequency vectors. Rather than integrating over the entire $p$-dimensional frequency space, we propose a sparse random projection approach  and compare empirical characteristic functions over adjacent moving
windows. 
{Rather than aiming to detect every possible distributional change in the full
$p$-dimensional space, we focus on changes that admit an informative
low-dimensional projection, thus restricting attention to a region that is both statistically meaningful
and computationally feasible in high dimensions. Sparse projections help mitigate noise
accumulation from irrelevant coordinates while avoiding an exhaustive search
over the full projection space. We further show that, whenever such an informative sparse direction exists, a
number of random projections growing only polynomially with $p$ is sufficient
to encounter one with high probability.}
{In addition, compared with other existing methods, the selected sparse projections provide low-dimensional representations of the detected distributional changes, facilitating interpretation of their nature and the coordinates potentially involved, as further illustrated in Section~\ref{sec:realdata}.} 

To accommodate nonstationary temporal dependence, we develop a Gaussian multiplier
bootstrap based on contrasts of local empirical characteristic functions. The local
construction allows the bootstrap procedure to adapt to the complex  dependence structure
without imposing  piecewise stationarity or
piecewise local stationarity \citep{zhou2013heteroscedasticity}. A key technical contribution, which may be of
independent interest, is a new Gaussian approximation for high-dimensional
multi-scale sparse time series under a general nonstationary dependence framework. The developed Gaussian approximation theory allows different coordinates
to have heterogeneous effective sample sizes, possibly of smaller order than
the full sample size $n$. Moreover, unlike existing Gaussian approximation
results \citep{chernozhukov2013gaussian,zhang2017gaussian,Zhang2018}, our
result does not require the coordinatewise variances to be uniformly bounded
away from zero. Such a condition is generally untenable for the  change-point statistics considered here, as some projected characteristic-function contrasts may have vanishing variances even after normalization by their effective sample sizes.

Finally, we embed the proposed bootstrap test into a recursive multiple
change point procedure. We establish the validity of the bootstrap under a
general nonstationary framework and provide asymptotic guarantees for estimating both the
number and locations of multiple change points. The resulting uniform
localization rate is close to the canonical rate up to logarithmic factors. To the best of
our knowledge, this is the first computationally tractable procedure for
testing and detecting general distributional changes in high dimensions while
allowing for the complex temporal dynamics.

The remainder of the paper is organized as follows. Section~\ref{sec:method} presents the
proposed methodology, and Section~\ref{sec:theory} establishes
the theoretical properties. Section~\ref{sec:simulation} reports
simulation studies, and Section~\ref{sec:realdata} presents a real-data application.  Technical details are deferred to the Supplementary
Material.

\section{Methodology}\label{sec:method}

For a probability law $\P$ on
$\mathbb R^p$, write
$\varphi_\P(\bm s)
    =
    \int_{\mathbb R^p}\exp(\i\bm s^\top\bm x)\,d\P(\bm x)$ 
for its characteristic function. Since a probability distribution is
uniquely determined by its characteristic function, $\P\neq \mathbb Q$ implies that
$\varphi_\P(\bm s)\neq\varphi_{\mathbb Q}(\bm s)$ for some $\bm s\in\mathbb R^p$. Empirical characteristic functions have been used to detect dependence 
\citep{Hong1999,Hong2001} and structural breaks 
\citep{FuHongWang2023} in time series.

In high dimensions, however, an exhaustive search for projection vectors over the full $p$-dimensional space $\mathbb R^p$ is
computationally infeasible. To address this issue,
 our proposed procedure exploits this characterization through $k_0$-sparse projection
vectors for some fixed $1\leq k_0\leq p$. In particular, if the marginal distributions of $\P$ and $\mathbb Q$ on a
coordinate set $\mathcal I\subseteq[p]$ with $|\mathcal I|\leq k_0$ are different, then there
exists a vector $\bm s$ supported on $\mathcal I$ such that
$\varphi_\P(\bm s)\neq\varphi_{\mathbb Q}(\bm s)$. 
The detailed detectability condition is further formalized
 in Section~\ref{sec:projection-complexity}. 


\begin{algorithm}[h!]
\caption{$k_0$-sparse random projection}
\label{alg:randomprojection}
\begin{algorithmic}[1]
\REQUIRE $k_0,R_0,J$
\ENSURE $\mathcal J=\{\bm s_1,\ldots,\bm s_J\}$

\STATE Initialize $\mathcal J\gets\emptyset$

\FOR{$j=1,\ldots,J$}
    \STATE Independently sample a support set
    $\mathcal I_j\sim
    \operatorname{Unif}
    \left(
    \left\{
    \mathcal I\subseteq[p]:
    |\mathcal I|=k_0
    \right\}
    \right)$.

    \STATE Conditional on $\mathcal I_j$, independently sample
    $\bm U_j\sim\operatorname{Unif}(\mathbb S^{k_0-1})$
    and $R_j\sim\operatorname{Unif}(0,R_0)$.

    \STATE Set
    $(\bm s_j)_{\mathcal I_j}=R_j\bm U_j$
    and $(\bm s_j)_{\mathcal I_j^c}=0$.

    \STATE Update
    $\mathcal J\gets\mathcal J\cup\{\bm s_j\}$.
\ENDFOR
\RETURN $\mathcal J$
\end{algorithmic}
\end{algorithm}


Let $\mathcal J=\{\bm s_1,\ldots,\bm s_J\}$ be the collection of sparse
frequency vectors generated by Algorithm~\ref{alg:randomprojection}, which is independent
of the observations $\{X_i\}_{i=1}^n$. 
For each $\bm s_j$, the
empirical characteristic functions computed over two adjacent windows are
compared through a moving-sum contrast. We then maximize the contrast jointly
over the candidate locations and the randomly generated sparse projections.
Specifically, for some index set $\mathcal B$, some window size $m$ and  bootstrap bandwidth $h$ with $m>h\geq 1$, 
\begin{align}\label{Tn(A1)}
    T_n(\mathcal B) = \max_{t\in \mathcal B} \max_{1\leq j\leq  J}\frac{1}{\sqrt{2m}}\left| \sum_{i=t-m+1}^{t}\exp\left(\i \bm s_j^\top \bm X_i\right)-  \sum_{i=t+1}^{t+m }\exp\left(\i \bm s_j^\top \bm X_i\right)\right|.
\end{align}
For a fixed $\bm s_j$, the two sums in $T_n(\mathcal B)$ are the
unnormalized empirical characteristic functions computed from the left and
right moving windows. Under the absence of a distributional change, their expectations coincide, while under   alternatives, the expectation of the contrast is nonzero
whenever the neighboring segment distributions have different
characteristic functions along $\bm s_j$. Taking the maximum over
$j=1,\ldots,J$ therefore allows the procedure to adapt to an unknown type and an unknown  support of the distributional change.


Theoretically, many classical omnibus discrepancies,
including the   energy distance, admit a weighted integral
representation involving the squared difference between two characteristic
functions over $\mathbb R^p$. Although such discrepancies characterize
equality of distributions when $p$ is fixed, their high-dimensional
expansions may be dominated by discrepancies in low-order moments. For instance, the Euclidean energy distance  asymptotically reduces to
comparisons of the means and the traces of the covariance matrices when $p$ is diverging, and
related kernel statistics exhibit analogous  phenomena in high dimensions
\citep{chakraborty2021framework,zhu2021interpoint,
yan2023kernel}.
Besides, a dense projection vector may also dilute a change that is concentrated on a
small subset of coordinates; see Remark \ref{rem:dense-projection}. By contrast, a
$k_0$-sparse projection vector performs a genuinely low-dimensional
distributional comparison, and avoids accumulating noise from unaffected
coordinates. 


We next develop a multiplier bootstrap procedure that asymptotically remains valid under
complex nonstationary  temporal  dependence. 
Define 
\begin{align*}
    \mathcal S_{j,t }^{(b)} = \frac{1}{\sqrt{2m }}\left(\sum_{l=t-m+1}^{t }Y_{jl}G_l^{(b)}- \sum_{l=t+1}^{t+m }Y_{jl}G_l^{(b)} \right),
\end{align*}
where  $\{G_l^{(b)}:1\leq l\leq n,\ 1\leq b\leq B\}$ are independent
standard Gaussian multipliers that are independent of the observations, and \begin{align*}
    Y_{jl} = \frac{1}{\sqrt{2h}}\left\{ \sum_{i=l-h+1}^{l}\exp\left(\i \bm s_j^\top\bm X_i\right) - \sum_{i=l +1}^{l+h}\exp\left(\i \bm s_j^\top\bm X_i\right) \right\}.
\end{align*}
The final bootstrap statistic is defined by  $\mathcal T_{n}^{(b)}(\mathcal B)=\max_{t\in \mathcal B} \max_{1\leq j\leq J} \left|\mathcal S_{j,t }^{(b)}\right|$, $1\leq b\leq B$.
The  local contrast $Y_{jl}$ is centered when $l$ is  away from the true change points, and multiplying these local
contrasts by independent Gaussian multipliers recovers the covariance
structure of the moving-sum statistic in (\ref{Tn(A1)})  without requiring stationarity or the estimation of a time-varying long-run covariance matrix. 

Let
\(
\widehat c_{1-\alpha}(\mathcal B)
:=
\inf\left\{
x:
\mathbb P\left(\mathcal T_{n}^{(b)}(\mathcal B)\le x\mid \bm X,\mathcal J\right)
\ge 1-\alpha
\right\}
\)
denote the conditional bootstrap $(1-\alpha)$-quantile. Since the bootstrap
approximation established in Theorem \ref{THM:H0} is valid away from a shrinking
neighborhood of zero, we use the slightly truncated critical value $\widetilde c_{1-\alpha}(\mathcal B)
=
\widehat c_{1-\alpha}(\mathcal B)\vee \rho_n$, where $\rho_n\asymp (\log\log n)^{-1}$. This truncation is introduced only for theoretical purposes and has negligible impact in practice. Let $\mathcal B_1=\left\{ m+h-1,m+h ,\ldots,n-m-h \right\}$. Then, $H_0$ is rejected when $T_n(\mathcal B_1)>\widetilde c_{1-\alpha}(\mathcal B_1)$.

To detect multiple change points, we recursively apply the bootstrap test
to the current candidate set. Whenever the null hypothesis of no remaining
change point on $\mathcal B$ is rejected, we record the location that maximizes the
moving-sum statistic 
and remove its $(1+\eta)m$-neighborhood from the
candidate set, where $\eta\in(0,1)$ is a fixed deletion parameter. The procedure
is repeated until the bootstrap test no longer rejects. The complete multiple change point 
detection procedure is summarized in Algorithm \ref{alg:cpdetection} below.


\begin{algorithm}[h!]
\caption{Change point detection}
\label{alg:cpdetection}
\begin{algorithmic}[1]
    \REQUIRE Initialize $\mathcal C\gets\varnothing$,
    $\mathcal B_1=\left\{m+h-1,m+h,\ldots,n-m-h\right\}$,
    $k=1$,   $\eta\in(0,1)$ and $\mathcal J=\{\bm s_1,\ldots,\bm s_J\}$.
    \ENSURE $\mathcal C$ contains the estimated change points.

    \WHILE{$\mathcal B_k\neq\varnothing$}
        \STATE Compute $T_n(\mathcal B_k)$ and
        $\widetilde c_{1-\alpha}(\mathcal B_k)$.
        \IF{$T_n(\mathcal B_k)\leq\widetilde c_{1-\alpha}(\mathcal B_k)$}
            \STATE \textbf{break}
        \ENDIF

        \STATE  $\widehat\tau_k
          =\arg\max_{t\in\mathcal B_k}\max_{1\leq j\leq J}
          \left|
            \sum_{i=t-m+1}^{t}
              \exp\!\left(\mathrm{i}\,\bm s_j^\top\bm X_i\right)-
            \sum_{i=t+1}^{t+m}
              \exp\!\left(\mathrm{i}\,\bm s_j^\top\bm X_i\right)
          \right|$. 

        \STATE Update $\mathcal C\gets
        \mathcal C\cup\{\widehat\tau_k\}$.
        \STATE Update $\mathcal B_{k+1}\gets
        \mathcal B_k\setminus
        \left[\widehat\tau_k-(1+\eta)m,\,
              \widehat\tau_k+(1+\eta)m\right]$.
        \STATE Set $k\gets k+1$.
    \ENDWHILE
    \RETURN $\mathcal C$.
\end{algorithmic}
\end{algorithm}

\section{Theoretical justification} 
\label{sec:theory}
This section establishes the theoretical properties of the proposed
procedures through three components: calibrating the maximum
of the resulting moving-sum processes, quantifying the computational complexity
of identifying an informative sparse projection vector,  and providing asymptotic guarantees  for estimating both the number and locations
of  change points. Throughout the theoretical analysis, $k_0$ in Algorithm \ref{alg:randomprojection} is fixed and
the projection radius $R_0$ is uniformly bounded above.

\subsection{Bootstrap consistency}

Throughout the paper, the $\mathcal L_q$ norm of a   random variable $X$ is
defined as $\left\|X\right\|_{q}=\left\{\E|X|^q\right\}^{1/q}$.  We next establish the asymptotic validity of the proposed multiplier bootstrap  under a general nonstationary physical
dependence framework \citep{wu2005nonlinear}. 
 For $i=1,\ldots,n$, assume that $\bm X_i=\bm{\mathcal G}_i(\cdots,e_{i-1},e_i)\in\mathbb R^p$, where $e_i$, $i\in\mathbb Z$, are i.i.d. random elements and $ \bm{\mathcal G}_i=(\mathcal G_{i,1},\ldots,\mathcal G_{i,p})^\top$ is a $p$-dimensional vector-valued Borel-measurable function. Notably, the physical mechanism process $\bm{\mathcal G}_i$ is allowed to vary with different $i$,   and neither strict stationarity nor (piecewise) local stationarity structure is imposed.  For $l\geq 0$ and $q\geq 1$, define the physical dependence measures of $\left\{\bm X_i\right\}_{i=1}^{n}$ by $\theta_{j,l,q}=\max_{1\leq i\leq n}\left\| \mathcal {  G}_{i,j}(\mathcal F_{i })- \mathcal {  G}_{i,j}(\mathcal F_{i,i-l })\right\|_q,$
    where $\mathcal F_{i }=(\cdots,  e_{i-l},e_{i-l+1},\cdots,e_{i-1},e_i)$, $\mathcal F_{i,i-l}=(\cdots,  e_{i-l}',e_{i-l+1},\cdots,e_{i-1},e_i)$, and  $ e_{i-l}'$ is an i.i.d. copy of $e_{i-l}$. 

Since $\mathcal J$
is generated independently of the observations, the  
vectors $\{\bm s_j\}_{j=1}^{J}$ can be treated as fixed throughout the
subsequent analysis. 
To investigate the null distribution of $T_n(\mathcal B)$, where $\mathcal B\subseteq \mathcal B_1$, 
define $\varepsilon_{ij}
    =
    \exp\left(i\bm s_j^\top\bm X_i\right)
    -
    \mathbb E\left\{\exp\left(i\bm s_j^\top\bm X_i\right)|\mathcal J\right\}$ for $1\leq i\leq n$, $1\leq j\leq J$.
We note that $T_n(\mathcal B)$ admits a useful high-dimensional representation:
it can be written as the $\ell_\infty$-norm of a normalized sum of $n$ sparse 
vectors. Formally, let $d=J|\mathcal B|$, and fix an arbitrary one-to-one correspondence $k\longleftrightarrow(t_k,j_k)
    \in\mathcal B\times\{1,\ldots,J\}$ for $1\leq k\leq d$. 
For each $1\leq i\leq n$ and $1\leq k\leq d$, define
\(
    x_{ik}
    =
    \left\{
        \bm 1(t_k-m+1\leq i\leq t_k)
        -
        \bm 1(t_k+1\leq i\leq t_k+m)
    \right\}
    \varepsilon_{i j_k}.
\)
Under $H_0$, the marginal distribution of $\bm X_i$ is constant over
time, so the deterministic components of the two adjacent windows cancel. 
Then, we rewrite 
\(
    T_n(\mathcal B)
    =
    \max_{1\leq k\leq d}
    \left|
 \sum_{i=1}^n x_{ik}
    \right|/\sqrt{2m}.
\)
Thus, the test statistic $T_n(\mathcal B)$ is the maximum of the moduli of normalized sums
of the coordinates of the $d$-dimensional  time series
\(
    \bm x_i=(x_{i1},\ldots,x_{id})^\top\in\mathbb C^d.
\)
For each coordinate $k$,  the sequence
$\{x_{ik}\}_{i=1}^n$ only contains exactly $2m$ potentially nonzero terms almost surely. 
The natural normalization is therefore $(2m)^{-1/2}$, rather than the
conventional normalization  $n^{-1/2}$. However, existing
 Gaussian approximation results for high-dimensional nonstationary time
series \citep{Zhang2018} are typically formulated for statistics scaled by $\sqrt n$, 
where each coordinate accumulates information over the full sample.
Since $m/n\to0$, normalization by $\sqrt n$ would make each sparse
coordinate asymptotically degenerate. Moreover, one cannot simply discard the time points corresponding to zero entries, since the locations of zero entries differ across coordinates.

To address this theoretical difficulty, Theorem S.1 in the
Supplementary Material develops a more general Gaussian approximation
theory for high-dimensional multi-scale sparse  nonstationary time series, which is of independent interest.
The developed framework allows different coordinates to contain different 
numbers of nonzero observations and to be normalized by their respective
effective sample sizes. Another theoretical contribution is the removal of the uniform lower-bound
condition on the coordinatewise variances, which is commonly imposed in
existing high-dimensional Gaussian approximation results \citep{chernozhukov2013gaussian,zhang2017gaussian,Zhang2018}. 

In the present setting, the effective sample size is $2m$ for every
coordinate. By further adopting the  simple convex set approximation technique in 
\citet{chernozhukov2017central,wu2024frequency}, the developed  Gaussian approximation
 for real-valued multi-scale sparse time series can be extended
to the complex-valued test statistic $T_n(\mathcal B)$, which further yields Theorem \ref{THM:H0} stated below.


\begin{itemize}

    \item[(A1)] For some $q> 4$ and $\chi\in(0,1)$,  $\max_{1\leq j\leq p}\theta_{j,l,q}=\O\left(\chi^l\right)$. 
    \item[(A2)] For some $b\in(0,3q/2-1)$, the number of projections $J\to\infty$, $J=\O\left(n^b\right)$. 
    \item[(A3)]   For some $\varpi\in(0,1]$, the minimum spacing $\vartheta_n\asymp n^{\varpi}$. 
    
   \item[(A4)] For some $\omega\in(0,\varpi)$, $\beta\in(0,\omega-4(1+2b)/q)$, the window size $m\asymp n^{\omega}$, the bootstrap bandwidth  $h\asymp n^{\beta}$. 
\end{itemize}

Assumption~(A1) requires the
effect of a remote innovation to decay exponentially over time. Similar short-range dependence conditions based on
physical dependence measures can be found in 
\citet{dette2019change,bai2024detecting}. 
Assumption~(A2) allows the number of random projections to grow
polynomially with the sample size \(n\). The restriction
\(b<3q/2-1\) guarantees that the remainder
\(\O\left(n^{b+1-3q/2}\right)\) in Theorem~\ref{THM:H0} vanishes, which
is needed to control the approximation error incurred when a circle in the
complex plane is approximated from both inside and outside by regular convex 
polygons. 
Assumptions~(A3)-(A4) specify the relative scales of the minimum
spacing between consecutive change points, the moving-sum window size \(m\),
and the bootstrap bandwidth \(h\). The parameter region in (A4) is nonempty  when
$q>4(1+2b)/\omega$.
\begin{theorem}\label{THM:H0}
Suppose that Assumptions (A1)-(A4) hold. Let $\iota_n$ be a positive sequence  of real numbers which diverges to infinity at an arbitrarily slow rate. Under $H_0$, there exists a sequence of
events $\{E_n\}_{n\geq 1}$ such that $\P(E_n)\to 1$, and on $E_n$, for any deterministic set $\mathcal B\subseteq \mathcal B_1$ and some constant $\delta>0$, \begin{align*}
    & \sup_{x\gtrsim (\log n)^{-\delta}}\left|\P\left(T_n(\mathcal B)\leq x|\mathcal J\right)-\P\left(\mathcal {T}_n^{(b)}(\mathcal B) \leq x|\bm X,\mathcal J\right)\right| \\\lesssim  &\left(J^{4/q}n^{2/q}\sqrt{h/m}+ 1/h  \right)^{1/3}\iota_n^{1/3}(\log n)^{1+2\delta} +m^{-1/8}(\log n )^{15/8+\delta}+n^{b+1-3q/2}.
\end{align*}
\end{theorem}

Theorem~\ref{THM:H0} establishes the validity of the proposed
multiplier bootstrap uniformly over all deterministic candidate sets
\(\mathcal B\subseteq \mathcal B_1\). The approximation is conditional on the realized projection
set $\mathcal J$, and the high-probability event is defined
under the joint law of the observations and the projection
randomization. 
The restriction $x\gtrsim(\log n)^{-\delta}$ arises from the absence
of a uniform lower bound on the coordinatewise variances. It is sufficient for our testing procedure because the truncated bootstrap
critical value satisfies
$\widetilde c_{1-\alpha}(\mathcal B)\ge\rho_n
\asymp(\log\log n)^{-1}$, and
$\rho_n\gg(\log n)^{-\delta}$ for every fixed $\delta>0$.


\subsection{Projection complexity}\label{sec:projection-complexity}

To quantify the computational budget required by the randomized projection
scheme, define the jump of characteristic functions  at the $\ell$th change
point by
\(
    g_\ell(\bm s)
    :=
    \mathbb E\exp\left(i\bm s^\top\bm X_{\tau_\ell+1}\right)
    -
    \mathbb E\exp\left(i\bm s^\top\bm X_{\tau_\ell}\right),\) \(\bm s\in\mathbb R^p.
\)
Define \(
    k^*:=\max_{1\leq\ell\leq r_n}k_\ell^*
\), 
\[
    k_\ell^*
    :=
    \min\left\{
        k\in\{1,\ldots,p\}:
        \exists\,\bm s\in\mathbb R^p
        \text{ s.t. }
        \|\bm s\|_0\leq k, \|\bm s\|_2\leq R_0,
        \text{ and }
        |g_\ell(\bm s)|>0
    \right\}.
\]
Under the alternative $H_1$, the condition imposed on $R_0$ in
Assumption~(B1) below ensures that each change point is detectable by
Algorithm~\ref{alg:randomprojection}, so that $k_\ell^*$ and $k^*$ are well defined.
Here, $k_\ell^*$ is the smallest sparsity level at which the $\ell$th
change point is detectable, and $k^*$ is the smallest common sparsity
level such that every change point is detectable.

Let $\bm s_\ell^\star$ denote the $k^*$-sparse informative projection
direction for the $\ell$th change point, namely,
\(
    \bm s_\ell^\star
    \in
    {\arg\max}_{\left\|\bm s\right\|_0\leq k^*,\left\|\bm s\right\|_2\leq R_0}
    |g_\ell(\bm s)|.
\)
The existence of \(\bm s_\ell^\star\) follows from continuity of \(g_\ell(\cdot)\) and compactness of the feasible set.
We note that the informative projection
directions need not be the same across different change points.
Let
\(
    \Delta_n
    :=
    \min_{1\leq\ell\leq r_n}
    |g_\ell(\bm s_\ell^\star)|
\) denote the minimum signal strength, evaluated at the informative projection direction,
over all change points. 
Before we proceed to investigate the property of Algorithm \ref{alg:randomprojection}, some mild assumptions are  introduced. 
\begin{itemize}
    \item[(B1)] 
 The inputs $R_0$ and  $k_0$ of Algorithm \ref{alg:randomprojection} satisfy that $\max_{||\bm s||_2\leq R_0}\left|g_\ell(\bm s)\right|>0$ for every $1\leq \ell\leq r_n$    and  $k_0 \geq k^\ast$.  
 \item[(B2)] For  some $C_{\mathbf\Sigma}>0$  possibly diverging with   $p$,  $\max_{0\leq \ell\leq r_n} \left\|\E \bm X_{\tau_\ell+1}\bm X_{\tau_\ell+1}^\top\right\|_{\op} \leq C_{\mathbf\Sigma}$.
\end{itemize}
\begin{theorem}\label{THM:computation_complexity}
    Suppose that $H_1$ holds. 
    Under Assumptions (B1)-(B2), for any small $\epsilon>0$ and any $c>0$,  \begin{align*}
    \P\left\{\min_{1\leq \ell\leq r_n}\max_{1\leq j\leq J}\left|g_\ell({\bm s_j})\right|>(1-\epsilon)\Delta_n\right\}>1-\exp(-c),
\end{align*} 
provided that the input parameter $J$  
in Algorithm \ref{alg:randomprojection} satisfies   $J
\geq
(c+\log r_n)
C_{k_0}^{-1}
\epsilon^{-k_0}
C_{\mathbf\Sigma}^{k_0/2}
p^{k^\ast}$, 
where  $C_{k_0}=\Delta_n/(8R_0)$ for $k_0=1$ and 
\[
C_{k_0}
=
\dfrac{k_0!\Gamma(k_0/2)}
{2\sqrt{\pi}(k_0-1)(k_0-k^*)!\Gamma((k_0-1)/2)}
\left(\dfrac{\Delta_n}{4R_0}\right)^{k_0},
 \quad k_0\ge 2.
\]
\end{theorem}

Theorem~\ref{THM:computation_complexity} shows that,  when $k_0$, $k^*$, $R_0$, and $C_{\mathbf\Sigma}$ 
 are bounded, and $\Delta_n$ is bounded away from zero, the  projection budget
\(
    J\gtrsim  p^{k^*}\log r_n
\) is sufficient: with high probability, every change point is associated
with at least one   projection direction in $\mathcal J$ produced by Algorithm \ref{alg:randomprojection}, along which the characteristic function
contrast approaches the minimum signal strength $\Delta_n$.  
Note that the exponent of the   dimension $p$ is governed
by the intrinsic sparsity level $k^*$. 
Thus, the required number of projections in Algorithm~\ref{alg:randomprojection}  grows only polynomially with $p$.

\begin{remark}\label{rem:dense-projection}
    A typical way in the literature to select a projection direction is to generate it elementwise independently from the standard normal distribution; see \cite{lopes2011more,liu2023new,liu2024random}. However, under the   alternative $H_1$ that $\{\bm X_i\}_{i=1}^{n}$ is a sequence of independent random vectors, where $\bm X_1,\ldots, \bm X_{\tau}\stackrel{\text{i.i.d.}}{\sim} \mathcal N(\bm 0,\mathbf I_p)$, and $\bm X_{\tau+1},\ldots, \bm X_{n}\stackrel{\text{i.i.d.}}{\sim} \mathcal N(\bm e_1,\mathbf I_p)$ with $\bm e_1=(1,0,\ldots,0)^\top$, if $\{\bm{s}_j\}_{j=1}^{J}$ is drawn from the  $p$-dimensional standard normal distributions with elementwise independence and the projection budget $J$ is restricted to any polynomial order of $p$,   the signal strength $\max_{1\leq j\leq J}\left| \E \exp\left(\i \bm s_j^\top \bm X_{\tau+1}\right)-\E \exp\left(\i \bm s_j^\top \bm X_{\tau}\right) \right|$ converges to zero in probability exponentially fast as the dimension $p$ diverges. This example illustrates the inefficiency of  dense Gaussian frequencies   in the high-dimensional regime  and motivates the use of sparse projections. 


\end{remark}

\subsection{Asymptotic guarantees for change point detection}

Combining this bootstrap approximation with the projection complexity guarantee in Theorem~\ref{THM:computation_complexity} allows us to analyze the proposed change point detection procedure. Let
\(
\widehat \tau_{(1)}<\cdots<\widehat \tau_{(\widehat  r_n)}
\)
denote the ordered estimates. The next theorem establishes asymptotic false-detection control under the null
and simultaneous guarantees for the estimated number and locations of the
change points under the alternative.
 
\begin{theorem}\label{THM:power}
Under Assumptions (A1)-(A4), for any $\alpha\in(0,1)$, 
\begin{itemize}
        \item[(a)] if there is no change point, then 
            $\liminf_{n\to\infty}\P\left(\widehat r_n=0\right)\geq 1-\alpha$. 
       \item[(b)]  Further  suppose that Assumptions (B1)-(B2), $J\gtrsim   C_{\mathbf\Sigma}^{k_0/2} p^{k^*}\Delta_n^{-k_0}\{\log (r_n\vee e)+c_n\}$ for some $c_n\to\infty$, and $m \Delta_n/(h\sqrt{\log n})\to\infty$ hold.  
       If there are $r_n\geq 1$ jumps and, for some constant
$\kappa>0$,
\(
 {(\log n)^{4+\kappa}}/{(m\Delta_n^2)}
\to0,
\)
then  for every fixed $c\in(0,\kappa)$,
  \begin{align*}
            \liminf_{n\to\infty}\P\left(\left\{\widehat r_n=r_n\right\}\cap \left\{\max_{1\leq k\leq r_n}\left|\widehat \tau_{(k)}-\tau_k\right|\lesssim   \Delta_n^{-2}(\log n)^{1+c} \right\}\right)\geq  1-\alpha.
        \end{align*}

    \end{itemize}
\end{theorem}

The first assertion of Theorem \ref{THM:power} provides asymptotic control of false detections under the null.
When there is no change point, the probability that the
procedure reports at least one spurious change point is asymptotically controlled
by the nominal level $\alpha$, which is inherited from the asymptotic
validity of the multiplier bootstrap established in
Theorem~\ref{THM:H0}. 
The second assertion gives a simultaneous guarantee for multiple change point detection
and localization under the alternative. With asymptotic probability at least
$1-\alpha$, the procedure recovers the correct number of change points and the estimated change point locations satisfy
\(
    \max_{1\le k\le r_n}
    |\widehat{\tau}_{(k)}-\tau_{k}|
    \lesssim
   \Delta_n^{-2}(\log n)^{1+c}.
\)
Thus, the uniform localization rate is close to the canonical
$\Delta_n^{-2}$ scale, up to a logarithmic factor whose exponent
can be arbitrarily close to one.

Assumption~(A2), together with the projection complexity requirement in
Theorem~\ref{THM:computation_complexity},   characterizes the dimensional
regime allowed by our theory. In particular, when $\Delta_n$ and
$C_{\mathbf\Sigma}$ are bounded away from zero and infinity, respectively, the dimension may diverge polynomially with the sample size, satisfying
\(
    p
    \ll
    n^{(q\omega-4)/(8k^*)},
\)
up to a logarithmic factor. 
This also highlights that the admissible dimensionality is governed by the  sparsity level $k^*$. 
Notably, although the moment parameter $q$ determines the admissible growth rate of the dimension $p$, it does not  enter the uniform localization rate itself.
This favorable feature is largely due to the boundedness of the
characteristic function, 
since
\(
    \left|\exp\left(\mathrm{i}\bm s^\top \bm X_i\right)\right|\equiv 1,
\)
regardless of the tail behavior of $\bm X_i$. 


\section{Simulations}\label{sec:simulation}
In this section, we evaluate the finite-sample performance of the proposed method through extensive simulation studies.

\subsection{Data-generating mechanisms}
\label{Subsection: Data-generating mechanisms}
Let $\bm Z_t\in\mathbb R^p$ be a latent  vector generated by
  $\bm Z_t = a_t \bm Z_{t-1} + \sqrt{1-a_t^2}\, \bm \varepsilon_t, \, 
  \bm \varepsilon_t \stackrel{\mathrm{iid}}{\sim} N(\bm 0, \mathbf I_p), \, 
  \bm Z_0 \sim N(\bm 0, \mathbf I_p)$.
For $t=1,\dots,n$, we consider three dependence mechanisms:  
$\text{IID:}\,   a_t\equiv 0,\, 
\text{LS:}\,a_t=0.5+0.3\sin(2\pi t/n),\,
\text{NS:}\,a_t = 0.5 + 0.2(-1)^t$.
The three DGPs have the same marginal distribution but different temporal dependence. The IID case provides an independent benchmark, while the locally stationary case has smoothly time-varying dependence. Under the nonstationary case, the lag-one dependence alternates between two values.

For each DGP, we consider $(n,p)\in\{(400,80),(800,160)\}$ under the alternatives and $n\in \{ 400, 800, 1200, 1600, 2000\}$ with $p=0.2n$ under the null. Under the alternatives, the affected coordinates are $\mathcal{I}=\{1,\ldots,s\}$, where $ s=\lfloor0.6p \rfloor$, and the coordinates in $[p]\setminus \mathcal{I}$ remain unchanged. We consider three hypotheses: (1) no change point; (2) a single change point occurs at $\tau_1=\lfloor n/2\rfloor$; (3) three change points occur at $(\tau_1,\tau_2,\tau_3) = \bigl(\lfloor n/4\rfloor, \lfloor n/2\rfloor, \lfloor 3n/4\rfloor\bigr)$. Let $r$ denote the number of change points. Set $\tau_0=0$ and $\tau_{r+1}=n$. For $\tau_k<t\leq\tau_{k+1}$, $k\in\{0,\ldots,r\}$ denotes the segment index. Under the null, we generate data as $\bm{X}_t=\bm Z_t$. Under each alternative, we generate $\bm{X}_{t, \mathcal{I}^c} = \bm{Z}_{t, \mathcal{I}^c}$ and transform $\bm{Z}_{t, \mathcal{I}}$ as follows.

\begin{enumerate}
\item \textit{Mean changes.} For $j\in \mathcal{I}$, generate $X_{t,j} = Z_{t,j} + \mu_k$. The affected coordinates have marginal distribution $N(\mu_k,1)$ in segment $k$.

\item \textit{Variance changes.} For $j\in \mathcal{I}$, generate $X_{t,j} = \sqrt{v_k}\,Z_{t,j}$. The affected coordinates have marginal distribution $N(0,v_k)$ in segment $k$.

\item \textit{Covariance changes.}  Let $\mathbf{\Sigma}_s(\rho) = (1-\rho)\mathbf{I}_s + \rho\,\bm{1}_s\bm{1}_s^\top$ be the $s\times s$ correlation matrix. If $\mathbf{L}_k$ satisfies $\mathbf{L}_k\mathbf{L}_k^\top=\mathbf{\Sigma}_k$, generate $\bm{X}_{t,\mathcal{I}} = \mathbf{L}_k \bm{Z}_{t, \mathcal{I}}$. The affected coordinates remain marginally $N(0,1)$, while their covariance matrix changes across segments.

\item \textit{Marginal distributional changes.}  Let $\Phi(\cdot)$ be the standard normal distribution function and let $U_{t,j}=\Phi(Z_{t,j})$. For $j\in \mathcal{I}$, generate $X_{t,j} = F_k^{-1}(U_{t,j})$. The four standardized marginal distributions are
  $\mathscr G=N(0,1), \,
  \mathscr R=\mathrm{Rademacher}\{-1,1 \}, \,
  \mathscr T=\sqrt{3/5}t_5, \,
  \mathscr U=\sqrt{3}\operatorname{Unif}(-1,1)$.
All four distributions have mean zero and variance one. Thus, the changes affect the marginal shape but not the first two moments.

\item \textit{Joint distributional changes.}
Let $s_1=\lfloor s/2\rfloor$, $V_t=Z_{t,1}$, and $W_t=\Phi^{-1}\{2\Phi(|V_t|)-1\}$. Since $2\Phi(|V_t|)-1$ is uniformly distributed on $(0,1)$, $W_t\sim N(0,1)$. For $\tau_k<t\le\tau_{k+1}$, generate
\[
 X_{t,1}=V_t,\qquad
 X_{t,j}=
 \begin{cases}
 \sqrt{\rho}\,V_t+\sqrt{1-\rho}\,Z_{t,j},
     &2\le j\le s_1,\\[2pt]
 \sqrt{\rho}\,b_kW_t+\sqrt{1-\rho}\,Z_{t,j},
     &s_1<j\le s,
 \end{cases}
 \qquad \rho=0.95,
\]
while $X_{t,I^c}=Z_{t,I^c}$. We use $(b_0,b_1)=(1,-1)$ for one change and $(b_0,b_1,b_2,b_3)=(1,-1,1,-1)$ for three changes. All margins remain $N(0,1)$. Since $W_t$ is an even function of $V_t$,  \(\mathbb{E}(V_tW_t)=0\), leaving the covariance matrix unchanged. However, the copula changes as the nonlinear dependence structure is reversed.

\end{enumerate}
Table~\ref{Table: segment-values} lists the segment values used in the five alternatives.
\begin{table}[h!]
\centering
\caption{Segment values used in the five alternatives.}
\label{Table: segment-values}
\small
\begin{tabular}{lll}
\toprule
Alternative & One change & Multiple changes \\
\midrule
Mean & $(0,1.5)$ & $(0,1.5,3,4.5)$ \\
Variance & $(1,4)$ & $(1,4,0.25,1)$ \\
Covariance & $(I_s,\Sigma_s(0.95))$ & $(I_s,\Sigma_s(0.95),I_s,\Sigma_s(0.95))$ \\
Marginal distribution & $(\mathscr G,\mathscr R)$ & $(\mathscr G,\mathscr R,\mathscr T,\mathscr U)$ \\
Joint distribution & $(1, -1)$ & $(1, -1, 1, -1)$ \\
\bottomrule
\end{tabular}%
\end{table}

\subsection{Implementation and evaluation}
\label{Subsection: Implementation and evaluation}
The proposed method is implemented with projection level $k_0=2$, nominal level $\alpha=0.05$, and local deletion parameter $\eta=0.25$. 
To select $h$, define $$\boldsymbol{\psi}_i = \bigl( \cos(s_1^\top X_i), \sin(s_1^\top X_i), \ldots, \cos(s_J^\top X_i),\sin(s_J^\top X_i) \bigr)^\top.$$  For a positive integer $r$, 
\(
D_{\ell a}(r) = (2r)^{-1/2} \left( \sum_{i=\ell-r+1}^{\ell}\psi_{ia} - \sum_{i=\ell+1}^{\ell+r}\psi_{ia} \right)\),  $r\leq\ell\leq n-r$,  $1\leq a\leq2J$.
We estimate $L_{a,n} =  \operatorname{Var}\left( n^{-1/2}\sum_{i=1}^{n}\psi_{ia}\,\middle|\,\mathcal J \right)$ by $\widehat L_{a,n}(r) = \sum_{\ell=r}^{n-r}D_{\ell a}^{\,2}(r) / (n-2r+1)$, where \(\psi_{ia}\) is the \(a\)-th coordinate of \(\boldsymbol{\psi}_i\). Following \cite{wu2024frequency}, we could use a plug-in AMSE criterion averaged over the $2J$ coordinates. Let $b_n=\lfloor n^{1/3}\rfloor$ and $\ell_n=\lfloor n^{1/6}\rfloor$, and define
\begin{align*}
\widehat C_{1a,n}^{\,2} &=
\frac{1}{(n-4b_n+1)b_n} \sum_{i=2b_n}^{n-2b_n}H_{ia}^{\,2}, \, H_{ia} = \frac{1}{\sqrt{2b_n}} \left\{ \sum_{\ell=i-b_n+1}^{i}D_{\ell a}^{\,2}(b_n) - \sum_{\ell=i+1}^{i+b_n}D_{\ell a}^{\,2}(b_n) \right\},\\
\widehat C_{2a,n} &= \frac{3}{n-3\ell_n+1} \sum_{u=1}^{n-3\ell_n+1} Q_{ua}(\ell_n) \left\{ Q_{u+\ell_n,a}(\ell_n) - Q_{u+2\ell_n,a}(\ell_n) \right\}, \, Q_{ua}(r)=\sum_{i=u}^{u+r-1}\psi_{ia}.
\end{align*}
Then, we select  
\(
  h
=
\max\left\{ \left\lfloor \left(
 {2\sum_{a=1}^{2J}\widehat C_{2a,n}^{\,2}}/
{\sum_{a=1}^{2J}\widehat C_{1a,n}^{\,2}}
\right)^{1/3}n^{1/3}\right\rfloor,1\right\}.
\)  
With $h$ fixed, we select $m$ using a BIC-type criterion adapted from \cite{wu2024frequency}. For each candidate $m$, let $0=\widehat\tau_0(m) < \widehat\tau_1(m) < \cdots < \widehat\tau_{\widehat r(m)}(m) < \widehat\tau_{\widehat r(m)+1}(m) = n$ denote the segmentation returned by Algorithm~\ref{alg:cpdetection}.  For $q=0,\ldots,\widehat r(m)$, define
\[
\overline{\boldsymbol{\psi}}_q(m) = \frac{1}{ \widehat\tau_{q+1}(m)-\widehat \tau_q(m) } \sum_{i=\widehat\tau_q(m)+1}^{\widehat\tau_{q+1}(m)} \boldsymbol{\psi}_i, \quad \operatorname{SSE}_{\psi}(m) = \sum_{q=0}^{\widehat r(m)} \sum_{i=\widehat\tau_q(m)+1}^{\widehat\tau_{q+1}(m)} \left\| \boldsymbol{\psi}_i- \overline{\boldsymbol{\psi}}_q(m) \right\|_2^2.
\]
We can select $m\in \left( \lfloor n^{16/29}\rfloor,\lfloor n^{2/3}\rfloor\right)$ as suggested to minimize
\[
\operatorname{BIC}_{\psi}(m) = n \log \left\{  {\operatorname{SSE}_{\psi}(m)}/{(2Jn)} \right\} + \widehat r(m)\log n.
\]
In practice, the choice of \(J\) and $R_0$ is typically guided by computational constraints. According to Theorem \ref{THM:computation_complexity}, the number of random projections required to capture sufficiently informative signals grows polynomially with the dimension \(p\). Motivated by this result, we empirically adopt a linear projection budget \(J=\lfloor 3.75p\rfloor+200\). As shown in Figure \ref{fig:ls-J-sensitivity}, further increases in \(J\) yield limited improvements in detection performance once \(J\) is sufficiently large, while incurring additional computational cost. Furthermore, the projection radius $R_0$ determines the frequency range explored by the procedure and depends on the scale of the observations. When the coordinates have substantially different scales, we recommend first standardizing each coordinate. While a sufficiently large $R_0$ is desirable for capturing informative frequency regions, Theorem \ref{THM:computation_complexity} suggests that an excessively large radius may increase the computational burden. Motivated by this trade-off and the sensitivity analysis in Section \ref{Subsection: Discussion on the parameters}, we adopt $R_0=3$ as a moderate default in our numerical experiments.
Critical values are computed using $B=200$ multiplier-bootstrap replications, and each reported value is based on $200$ Monte Carlo replications. 

We compare the proposed method with three representative distributional change point detection methods: gSeg \citep{chu2019asymptotic}, kerSeg \citep{song2022practical}, and UStat \citep{boniece2024change}. 
gSeg detects distributional changes by scanning edge counts on a similarity graph and we use a 5-nearest-neighbor graph, and kerSeg scans standardized kernel similarities between the subsamples before and after each candidate location. For kerSeg, we use a Gaussian kernel with bandwidth $\operatorname{median}_{i<j}\|\bm{X}_i-\bm{X}_j\|_2$, and set $r_1=1.2$ and $r_2=0.8$ defined therein, which control the relative weights assigned to kernel similarities before and after each candidate change point. UStat is based on pairwise $\ell_1$ distances and combines two U-statistic processes that are sensitive to different types of change. We set  $\beta=0.9$ and $\kappa=0.4$, where $\beta$ controls the balance between the two component processes and $\kappa$ controls the weight of candidate locations near the boundaries. All tuning parameters follow the recommendations in the corresponding papers. 

Note that the competing methods are originally designed for the at-most-one-change setting, with  multiple change point extensions via recursive procedures, such as binary segmentation (BS). Thus, we use their single-change procedures in our at-most-one-change experiments and their BS extensions, denoted by gSeg$^\star$, kerSeg$^\star$, and UStat$^\star$ respectively, in the multiple-change experiments. 

For evaluation metrics, $\bar{\hat r}$ is the average of the estimated number of change points. Under the null, \textit{Size} is the empirical probability of reporting at least one change point. Under the alternatives, an estimated change point is matched to a true change point if it lies within $3\log n$ time points, and each estimated change point can be matched at most once. \textit{All} is the empirical probability that all true change points are matched, and \textit{Perfect} further requires that there are no false detections. MAD is the mean absolute localization error over matched change points, and ARI is the adjusted Rand index between the true and estimated segmentations.

\subsection{Results} 
Table~\ref{Table: Null size}  reports the empirical size at the nominal level $\alpha=0.05$. The proposed method keeps the empirical size close to the nominal level across all DGPs and sample sizes, while the competing methods control the Type-I error only under the IID DGP. Under LS and NS, they reject every null sample and yield severe false detections. 
\begin{table}[h!]
\centering
\caption{Empirical size under the null at the nominal level $0.05$.}
\label{Table: Null size}
\small
\setlength{\tabcolsep}{3.2pt}
\begin{tabular}{llccccc}
\toprule
DGP & Methods
& $n=400$ & $n=800$ & $n=1200$ & $n=1600$ & $n=2000$\\
\midrule
\multirow{4}{*}{IID}
& Proposed &  \textbf {0.055}  &  \textbf{0.050} &  \textbf{0.040} &  \textbf{0.040} & \textbf{0.065}\\
& gSeg     & 0.055 & 0.085 & 0.060 & 0.085 & 0.055\\
& kerSeg   & 0.020 & 0.030 & 0.045 & 0.025 & 0.045\\
& UStat    & 0.070 & 0.055 & 0.075 & 0.050 & 0.055\\
\midrule
\multirow{4}{*}{LS}
& Proposed & \textbf{0.060} & \textbf{0.055} & \textbf{0.055} & \textbf{0.045} & \textbf{0.055}\\
& gSeg
& {1.000} &  {1.000}
& {1.000} & {1.000}
& {1.000}\\
& kerSeg
& {1.000} &  {1.000}
&  {1.000} &  {1.000}
& {1.000}\\
& UStat
& {1.000} & {1.000}
& {1.000} & {1.000}
& {1.000}\\
\midrule
\multirow{4}{*}{NS}
& Proposed & \textbf{0.045} & \textbf{0.055} & \textbf{0.055} & \textbf{0.045} & \textbf{0.050}\\
& gSeg
& {1.000} & {1.000}
& {1.000} & {1.000}
& {1.000}\\
& kerSeg
& {1.000} & {1.000}
& {1.000} & {1.000}
& {1.000}\\
& UStat
& {1.000} & {1.000}
& {1.000} & {1.000}
& {1.000}\\
\bottomrule
\end{tabular}
\end{table}

Tables~\ref{Table: One change IID}--\ref{Table: One change NS} report the performance under the one-change alternatives. The proposed method performs well across all change types and DGPs, reliably detecting the true change and yielding accurate segmentations. Even under the IID setting, the competing methods exhibit clear limitations beyond simple mean and variance shifts. For the marginal distribution change, although gSeg frequently rejects the null hypothesis, it often fails to localize the change point accurately, whereas kerSeg suffers from a low rejection rate. For the covariance change, UStat has limited power because of its pairwise $\ell_1$ distance criterion, $\mathbb{E}\|\bm{X}-\bm{Y}\|_1 = \sum_{j=1}^{p}\mathbb{E}|X_j-Y_j|$. For the joint distribution change, unchanged margins leave UStat without a population contrast in mean $\ell_1$ distances, while kerSeg appears less sensitive to this nonlinear dependence change.

The gap becomes larger under LS and NS. The values of $\bar{\hat r}$ near one for the competing methods do not indicate successful detection. For several alternatives, their low \textit{Perfect} rates show that the reported location is not close to the true change point. Moreover, their empirical sizes under the corresponding null models are close to one, suggesting that time-varying temporal dependence may be misinterpreted as evidence of a distributional change. Under the alternatives, this false signal may compete with or dominate the true change. In contrast, the proposed method maintains Type I error control and high \textit{Perfect} rates under different DGPs.

\begin{table}[h!]
\centering
\caption{One-change performance under the IID mechanism.}
\label{Table: One change IID}
\small
\setlength{\tabcolsep}{3.2pt}
\begin{tabular}{ll*{5}{c}@{\hspace{5pt}}*{5}{c}}
\toprule
&& \multicolumn{5}{c}{$n=400,\ p=80$}
& \multicolumn{5}{c}{$n=800,\ p=160$}\\
\cmidrule(lr){3-7}\cmidrule(lr){8-12}
Case & Method
& $\overline{\widehat r}$ & All & Perfect & MAD & ARI
& $\overline{\widehat r}$ & All & Perfect & MAD & ARI\\
\midrule
Mean
& Proposed
& 1.015 & 1.000 & 0.985 & 0.820 & 0.988
& 1.040 & 1.000 & 0.960 & 0.670 & 0.987\\
& gSeg
& 1.000 & 1.000 & 1.000 & 0.700 & 0.993
& 1.000 & 1.000 & 1.000 & 0.000 & 1.000\\
& kerSeg
& 1.000 & 1.000 & 1.000 & 0.440 & 0.996
& 1.000 & 1.000 & 1.000 & 0.450 & 0.998\\
& UStat
& 1.000 & 1.000 & 1.000 & 0.480 & 0.995
& 1.000 & 1.000 & 1.000 & 0.000 & 1.000\\
\addlinespace[2pt]
Variance
& Proposed
& 1.015 & 0.960 & 0.940 & 3.541 & 0.949
& 1.035 & 0.965 & 0.930 & 4.176 & 0.966\\
& gSeg
& 1.000 & 0.990 & 0.990 & 2.086 & 0.977
& 1.000 & 0.990 & 0.990 & 2.322 & 0.988\\
& kerSeg
& 1.000 & 1.000 & 1.000 & 0.035 & 1.000
& 1.000 & 1.000 & 1.000 & 0.005 & 1.000\\
& UStat
& 1.000 & 1.000 & 1.000 & 0.005 & 1.000
& 1.000 & 1.000 & 1.000 & 0.000 & 1.000\\
\addlinespace[2pt]
Covariance
& Proposed
& 1.035 & 0.995 & 0.960 & 1.814 & 0.972
& 1.065 & 1.000 & 0.935 & 1.655 & 0.976\\
& gSeg
& 1.000 & 0.910 & 0.910 & 3.302 & 0.940
& 1.000 & 0.925 & 0.925 & 4.378 & 0.970\\
& kerSeg
& 1.000 & 0.990 & 0.990 & 1.995 & 0.975
& 1.000 & 0.995 & 0.995 & 1.417 & 0.992\\
& UStat
& 0.215 & 0.010 & 0.010 & 6.500 & 0.077
& 0.220 & 0.010 & 0.010 & 13.000 & 0.083\\
\addlinespace[2pt]
Marginal
& Proposed
& 1.035 & 1.000 & 0.965 & 0.985 & 0.982
& 1.010 & 0.995 & 0.985 & 0.960 & 0.992\\
distribution
& gSeg
& 0.790 & 0.380 & 0.380 & 5.250 & 0.537
& 0.985 & 0.480 & 0.480 & 6.354 & 0.773\\
& kerSeg
& 0.075 & 0.000 & 0.000 & -- & 0.026
& 0.070 & 0.005 & 0.005 & 14.000 & 0.021\\
& UStat
& 1.000 & 1.000 & 1.000 & 0.110 & 0.999
& 1.000 & 1.000 & 1.000 & 0.005 & 1.000\\
\addlinespace[2pt]
Joint
& Proposed
& 1.030 & 0.980 & 0.955 & 2.362 & 0.966
& 1.055 & 0.990 & 0.935 & 2.697 & 0.972\\
distribution
& gSeg
& 1.000 & 1.000 & 1.000 & 0.325 & 0.997
& 1.000 & 1.000 & 1.000 & 0.031 & 0.998\\
& kerSeg
& 0.275 & 0.100 & 0.100 & 7.900 & 0.154
& 0.550 & 0.365 & 0.365 & 6.671 & 0.474\\
& UStat
& 0.055 & 0.005 & 0.005 & 17.000 & 0.020
& 0.055 & 0.000 & 0.000 & -- & 0.025\\
\bottomrule
\end{tabular}
\end{table}

\begin{table}[h!]
\centering
\caption{One-change performance under the LS mechanism.}
\label{Table: One change LS}
\small
\setlength{\tabcolsep}{3.5pt}
\begin{tabular}{ll*{5}{c}@{\hspace{6pt}}*{5}{c}}
\toprule
&& \multicolumn{5}{c}{$n=400,\ p=80$}
& \multicolumn{5}{c}{$n=800,\ p=160$}\\
\cmidrule(lr){3-7}\cmidrule(lr){8-12}
Case & Method
& $\overline{\widehat r}$ & All & Perfect & MAD & ARI
& $\overline{\widehat r}$ & All & Perfect & MAD & ARI\\
\midrule
Mean
& Proposed
& 1.000 & 1.000 & 1.000 & 1.430 & 0.986
& 1.010 & 1.000 & 0.990 & 1.770 & 0.989\\
& gSeg
& 1.000 & 1.000 & 1.000 & 0.000 & 1.000
& 1.000 & 1.000 & 1.000 & 0.000 & 1.000\\
& kerSeg
& 1.000 & 1.000 & 1.000 & 0.210 & 0.998
& 1.000 & 1.000 & 1.000 & 0.260 & 0.999\\
& UStat
& 1.000 & 1.000 & 1.000 & 0.000 & 1.000
& 1.000 & 1.000 & 1.000 & 0.000 & 1.000\\
\addlinespace[2pt]
Variance
& Proposed
& 1.000 & 0.800 & 0.800 & 4.363 & 0.907
& 1.060 & 0.875 & 0.855 & 4.909 & 0.943\\
& gSeg
& 1.000 & 0.000 & 0.000 & -- & 0.042
& 1.000 & 0.000 & 0.000 & -- & 0.126\\
& kerSeg
& 1.000 & 1.000 & 1.000 & 0.075 & 0.999
& 1.000 & 1.000 & 1.000 & 0.010 & 1.000\\
& UStat
& 1.000 & 1.000 & 1.000 & 0.010 & 1.000
& 1.000 & 1.000 & 1.000 & 0.000 & 1.000\\
\addlinespace[2pt]
Covariance
& Proposed
& 1.000 & 0.990 & 0.990 & 2.061 & 0.978
& 1.005 & 0.990 & 0.990 & 2.253 & 0.987\\
& gSeg
& 1.000 & 0.000 & 0.000 & -- & 0.516
& 1.000 & 0.000 & 0.000 & -- & 0.676\\
& kerSeg
& 1.000 & 0.790 & 0.790 & 3.285 & 0.880
& 1.000 & 0.985 & 0.985 & 1.411 & 0.991\\
& UStat
& 0.935 & 0.125 & 0.125 & 8.560 & 0.377
& 0.955 & 0.070 & 0.070 & 8.643 & 0.407\\
\addlinespace[2pt]
Marginal
& Proposed
& 1.000 & 1.000 & 1.000 & 1.265 & 0.988
& 1.000 & 1.000 & 1.000 & 1.130 & 0.994\\
distribution
& gSeg
& 1.000 & 0.000 & 0.000 & -- & 0.138
& 1.000 & 0.000 & 0.000 & -- & 0.084\\
& kerSeg
& 1.000 & 0.005 & 0.005 & 6.000 & 0.337
& 1.000 & 0.000 & 0.000 & -- & 0.313\\
& UStat
& 1.000 & 0.995 & 0.995 & 0.236 & 0.997
& 1.000 & 1.000 & 1.000 & 0.015 & 1.000\\
\addlinespace[2pt]
Joint
& Proposed
& 1.000 & 0.955 & 0.955 & 3.508 & 0.955
& 1.030 & 0.960 & 0.940 & 3.698 & 0.971\\
distribution
& gSeg
& 1.000 & 1.000 & 1.000 & 0.805 & 0.992
& 1.000 & 1.000 & 1.000 & 0.425 & 0.998\\
& kerSeg
& 0.995 & 0.200 & 0.200 & 7.775 & 0.469
& 1.000 & 0.210 & 0.210 & 6.476 & 0.546\\
& UStat
& 0.575 & 0.070 & 0.070 & 9.571 & 0.243
& 0.610 & 0.045 & 0.045 & 9.111 & 0.236\\
\bottomrule
\end{tabular}
\end{table}

\begin{table}[h!]
\centering
\caption{One-change performance under the NS mechanism.}
\label{Table: One change NS}
\small
\setlength{\tabcolsep}{3.2pt}
\begin{tabular}{ll*{5}{c}@{\hspace{5pt}}*{5}{c}}
\toprule
&& \multicolumn{5}{c}{$n=400,\ p=80$}
& \multicolumn{5}{c}{$n=800,\ p=160$}\\
\cmidrule(lr){3-7}\cmidrule(lr){8-12}
Case & Method
& $\overline{\widehat r}$ & All & Perfect & MAD & ARI
& $\overline{\widehat r}$ & All & Perfect & MAD & ARI\\
\midrule
Mean
& Proposed
& 1.005 & 1.000 & 0.995 & 1.445 & 0.985
& 1.000 & 1.000 & 1.000 & 1.375 & 0.993\\
& gSeg
& 1.000 & 1.000 & 1.000 & 0.000 & 1.000
& 1.000 & 1.000 & 1.000 & 0.000 & 1.000\\
& kerSeg
& 1.000 & 1.000 & 1.000 & 0.120 & 0.999
& 1.000 & 1.000 & 1.000 & 0.180 & 0.999\\
& UStat
& 1.000 & 1.000 & 1.000 & 0.140 & 0.999
& 1.000 & 1.000 & 1.000 & 0.000 & 1.000\\
\addlinespace[2pt]
Variance
& Proposed
& 1.045 & 0.905 & 0.870 & 4.442 & 0.923
& 1.015 & 0.935 & 0.920 & 5.294 & 0.964\\
& gSeg
& 1.000 & 0.015 & 0.015 & 0.667 & 0.053
& 1.000 & 0.000 & 0.000 & -- & 0.022\\
& kerSeg
& 1.000 & 1.000 & 1.000 & 0.085 & 0.999
& 1.000 & 1.000 & 1.000 & 0.000 & 1.000\\
& UStat
& 1.000 & 1.000 & 1.000 & 0.010 & 1.000
& 1.000 & 1.000 & 1.000 & 0.000 & 1.000\\
\addlinespace[2pt]
Covariance
& Proposed
& 1.015 & 1.000 & 0.985 & 2.380 & 0.973
& 1.000 & 0.995 & 0.995 & 1.910 & 0.990\\
& gSeg
& 1.000 & 0.190 & 0.190 & 3.737 & 0.291
& 1.000 & 0.060 & 0.060 & 6.333 & 0.155\\
& kerSeg
& 1.000 & 0.755 & 0.755 & 3.093 & 0.849
& 1.000 & 0.960 & 0.960 & 2.760 & 0.974\\
& UStat
& 0.580 & 0.110 & 0.110 & 6.864 & 0.256
& 0.655 & 0.035 & 0.035 & 7.857 & 0.256\\
\addlinespace[2pt]
Marginal
& Proposed
& 1.000 & 1.000 & 1.000 & 1.160 & 0.988
& 1.000 & 1.000 & 1.000 & 1.325 & 0.993\\
distribution
& gSeg
& 1.000 & 0.000 & 0.000 & -- & 0.120
& 1.000 & 0.000 & 0.000 & -- & 0.070\\
& kerSeg
& 1.000 & 0.055 & 0.055 & 7.727 & 0.270
& 1.000 & 0.030 & 0.030 & 14.667 & 0.240\\
& UStat
& 1.000 & 1.000 & 1.000 & 0.230 & 0.998
& 1.000 & 1.000 & 1.000 & 0.000 & 1.000\\
\addlinespace[2pt]
Joint
& Proposed
& 1.015 & 0.955 & 0.940 & 3.581 & 0.953
& 1.010 & 0.990 & 0.985 & 3.823 & 0.978\\
distribution
& gSeg
& 1.000 & 1.000 & 1.000 & 0.715 & 0.993
& 1.000 & 1.000 & 1.000 & 0.390 & 0.998\\
& kerSeg
& 0.945 & 0.215 & 0.215 & 5.395 & 0.448
& 0.980 & 0.385 & 0.385 & 7.844 & 0.681\\
& UStat
& 0.300 & 0.035 & 0.035 & 10.143 & 0.120
& 0.295 & 0.020 & 0.020 & 9.000 & 0.118\\
\bottomrule
\end{tabular}
\end{table}

Tables \ref{Table: Three changes IID}--\ref{Table: Three changes NS}  report performance in the three-change settings. Each of the competing methods still fails to detect certain types of change points. Moreover, they report a large number of change points under LS and NS, and the overestimation becomes more severe as $n$ increases. This further illustrates that, under complex temporal dependence, procedures calibrated for i.i.d. observations may mistake changes in the temporal dynamics for  distributional changes,  resulting in high \textit{All} and near zero \textit{Perfect} rates. In contrast, the proposed method estimates about three changes with consistently high \textit{All} and \textit{Perfect} rates.

\begin{table}[h!]
\centering
\caption{Three-change performance under the IID mechanism.}
\label{Table: Three changes IID}
\small
\setlength{\tabcolsep}{3.0pt}
\begin{tabular}{ll*{5}{c}@{\hspace{5pt}}*{5}{c}}
\toprule
&& \multicolumn{5}{c}{$n=400,\ p=80$}
& \multicolumn{5}{c}{$n=800,\ p=160$}\\
\cmidrule(lr){3-7}\cmidrule(lr){8-12}
Case & Method
& $\overline{\widehat r}$ & All & Perfect & MAD & ARI
& $\overline{\widehat r}$ & All & Perfect & MAD & ARI\\
\midrule
Mean
& Proposed
& 3.000 & 1.000 & 1.000 & 0.778 & 0.985
& 3.000 & 1.000 & 1.000 & 0.710 & 0.993\\
& gSeg$^\star$
& 3.450 & 1.000 & 0.620 & 0.000 & 0.984
& 3.315 & 1.000 & 0.770 & 0.000 & 0.989\\
& kerSeg$^\star$
& 3.185 & 1.000 & 0.825 & 0.533 & 0.990
& 3.170 & 1.000 & 0.845 & 0.212 & 0.992\\
& UStat$^\star$
& 3.345 & 1.000 & 0.795 & 0.000 & 0.985
& 3.510 & 1.000 & 0.790 & 0.000 & 0.982\\
\addlinespace[2pt]
Variance
& Proposed
& 3.000 & 0.940 & 0.940 & 2.724 & 0.939
& 3.000 & 0.960 & 0.960 & 3.171 & 0.966\\
& gSeg$^\star$
& 4.560 & 1.000 & 0.150 & 0.657 & 0.955
& 5.295 & 1.000 & 0.080 & 0.748 & 0.958\\
& kerSeg$^\star$
& 3.210 & 1.000 & 0.830 & 0.103 & 0.988
& 3.240 & 1.000 & 0.805 & 0.027 & 0.988\\
& UStat$^\star$
& 3.360 & 1.000 & 0.800 & 0.013 & 0.983
& 3.265 & 1.000 & 0.830 & 0.000 & 0.987\\
\addlinespace[2pt]
Covariance
& Proposed
& 3.000 & 0.995 & 0.995 & 1.781 & 0.965
& 3.000 & 1.000 & 1.000 & 1.750 & 0.983\\
& gSeg$^\star$
& 4.835 & 0.985 & 0.140 & 1.114 & 0.908
& 5.535 & 1.000 & 0.065 & 1.802 & 0.930\\
& kerSeg$^\star$
& 1.995 & 0.250 & 0.065 & 1.981 & 0.413
& 3.620 & 0.795 & 0.265 & 1.333 & 0.821\\
& UStat$^\star$
& 0.830 & 0.020 & 0.000 & 3.850 & 0.067
& 1.150 & 0.015 & 0.000 & 4.615 & 0.073\\
\addlinespace[2pt]
Marginal
& Proposed
& 2.965 & 0.905 & 0.905 & 2.458 & 0.936
& 3.000 & 0.945 & 0.945 & 2.635 & 0.970\\
distribution
& gSeg$^\star$
& 0.335 & 0.000 & 0.000 & 4.917 & 0.072
& 1.095 & 0.075 & 0.030 & 5.523 & 0.258\\
& kerSeg$^\star$
& 0.090 & 0.000 & 0.000 & 4.400 & 0.014
& 0.050 & 0.000 & 0.000 & 15.000 & 0.008\\
& UStat$^\star$
& 5.220 & 0.770 & 0.255 & 1.542 & 0.865
& 5.575 & 0.865 & 0.370 & 1.393 & 0.915\\
\addlinespace[2pt]
Joint
& Proposed
& 3.000 & 0.995 & 0.995 & 2.482 & 0.951
& 3.000 & 0.985 & 0.985 & 2.446 & 0.975\\
distribution
& gSeg$^\star$
& 3.495 & 0.955 & 0.555 & 1.122 & 0.928
& 3.505 & 1.000 & 0.650 & 1.162 & 0.971\\
& kerSeg$^\star$
& 0.190 & 0.000 & 0.000 & 4.375 & 0.030
& 0.260 & 0.000 & 0.000 & 6.333 & 0.037\\
& UStat$^\star$
& 0.140 & 0.000 & 0.000 & 6.000 & 0.019
& 0.055 & 0.000 & 0.000 & 9.000 & 0.013\\
\bottomrule
\end{tabular}
\end{table}

\begin{table}[h!]
\centering
\caption{Three-change performance under the LS mechanism.}
\label{Table: Three changes LS}
\small
\setlength{\tabcolsep}{3.2pt}
\begin{tabular}{ll*{5}{c}@{\hspace{5pt}}*{5}{c}}
\toprule
&& \multicolumn{5}{c}{$n=400,\ p=80$}
& \multicolumn{5}{c}{$n=800,\ p=160$}\\
\cmidrule(lr){3-7}\cmidrule(lr){8-12}
Case & Method
& $\overline{\widehat r}$ & All & Perfect & MAD & ARI
& $\overline{\widehat r}$ & All & Perfect & MAD & ARI\\
\midrule
Mean
& Proposed
& 2.995 & 0.975 & 0.975 & 1.385 & 0.970
& 2.990 & 0.975 & 0.975 & 1.545 & 0.981\\
& gSeg$^\star$
& 51.425 & 1.000 & 0.000 & 0.000 & 0.131
& 109.710 & 1.000 & 0.000 & 0.000 & 0.056\\
& kerSeg$^\star$
& 57.480 & 1.000 & 0.000 & 0.000 & 0.112
& 122.325 & 1.000 & 0.000 & 0.000 & 0.047\\
& UStat$^\star$
& 70.680 & 1.000 & 0.000 & 0.000 & 0.181
& 166.415 & 1.000 & 0.000 & 0.000 & 0.062\\
\addlinespace[2pt]
Variance
& Proposed
& 2.985 & 0.850 & 0.850 & 3.883 & 0.906
& 3.000 & 0.860 & 0.860 & 3.886 & 0.950\\
& gSeg$^\star$
& 50.380 & 1.000 & 0.000 & 1.022 & 0.143
& 108.010 & 1.000 & 0.000 & 0.830 & 0.060\\
& kerSeg$^\star$
& 55.935 & 1.000 & 0.000 & 0.200 & 0.130
& 121.230 & 1.000 & 0.000 & 0.057 & 0.050\\
& UStat$^\star$
& 68.310 & 1.000 & 0.000 & 0.017 & 0.205
& 163.970 & 1.000 & 0.000 & 0.000 & 0.069\\
\addlinespace[2pt]
Covariance
& Proposed
& 2.965 & 0.945 & 0.945 & 2.215 & 0.943
& 3.000 & 0.985 & 0.985 & 2.087 & 0.978\\
& gSeg$^\star$
& 43.220 & 1.000 & 0.000 & 1.372 & 0.250
& 89.585 & 1.000 & 0.000 & 1.317 & 0.157\\
& kerSeg$^\star$
& 45.405 & 0.985 & 0.000 & 1.129 & 0.286
& 92.310 & 1.000 & 0.000 & 0.868 & 0.270\\
& UStat$^\star$
& 31.080 & 0.275 & 0.000 & 2.551 & 0.278
& 70.935 & 0.390 & 0.000 & 2.097 & 0.250\\
\addlinespace[2pt]
Marginal
& Proposed
& 2.925 & 0.870 & 0.870 & 2.345 & 0.929
& 3.000 & 0.925 & 0.925 & 2.362 & 0.970\\
distribution
& gSeg$^\star$
& 50.275 & 0.885 & 0.000 & 3.057 & 0.124
& 108.225 & 1.000 & 0.000 & 2.472 & 0.056\\
& kerSeg$^\star$
& 56.285 & 0.855 & 0.000 & 2.496 & 0.113
& 121.610 & 0.975 & 0.000 & 1.926 & 0.048\\
& UStat$^\star$
& 69.995 & 0.945 & 0.000 & 1.177 & 0.173
& 165.130 & 0.995 & 0.000 & 0.715 & 0.066\\
\addlinespace[2pt]
Joint
& Proposed
& 2.930 & 0.875 & 0.875 & 3.075 & 0.915
& 2.985 & 0.930 & 0.930 & 3.690 & 0.955\\
distribution
& gSeg$^\star$
& 29.145 & 1.000 & 0.000 & 0.987 & 0.475
& 59.230 & 1.000 & 0.000 & 0.842 & 0.352\\
& kerSeg$^\star$
& 27.670 & 0.195 & 0.000 & 3.403 & 0.359
& 53.265 & 0.255 & 0.000 & 3.565 & 0.368\\
& UStat$^\star$
& 6.110 & 0.015 & 0.000 & 4.828 & 0.174
& 11.740 & 0.010 & 0.000 & 4.538 & 0.183\\
\bottomrule
\end{tabular}
\end{table}

\begin{table}[h!]
\centering
\caption{Three-change performance under the NS mechanism.}
\label{Table: Three changes NS}
\small
\setlength{\tabcolsep}{2.8pt}
\begin{tabular}{ll*{5}{c}@{\hspace{4pt}}*{5}{c}}
\toprule
&& \multicolumn{5}{c}{$n=400,\ p=80$}
& \multicolumn{5}{c}{$n=800,\ p=160$}\\
\cmidrule(lr){3-7}\cmidrule(lr){8-12}
Case & Method
& $\overline{\widehat r}$ & All & Perfect & MAD & ARI
& $\overline{\widehat r}$ & All & Perfect & MAD & ARI\\
\midrule
Mean
& Proposed
& 3.000 & 0.995 & 0.995 & 1.215 & 0.976
& 3.000 & 1.000 & 1.000 & 1.527 & 0.985\\
& gSeg$^\star$
& 56.005 & 1.000 & 0.000 & 0.000 & 0.100
& 115.530 & 1.000 & 0.000 & 0.000 & 0.047\\
& kerSeg$^\star$
& 69.440 & 1.000 & 0.000 & 0.000 & 0.081
& 141.010 & 1.000 & 0.000 & 0.000 & 0.040\\
& UStat$^\star$
& 107.670 & 1.000 & 0.000 & 0.000 & 0.043
& 225.140 & 1.000 & 0.000 & 0.000 & 0.020\\
\addlinespace[2pt]
Variance
& Proposed
& 3.000 & 0.940 & 0.940 & 3.397 & 0.927
& 3.000 & 0.940 & 0.940 & 3.588 & 0.960\\
& gSeg$^\star$
& 55.235 & 1.000 & 0.000 & 0.997 & 0.101
& 115.060 & 1.000 & 0.000 & 0.655 & 0.048\\
& kerSeg$^\star$
& 69.325 & 1.000 & 0.000 & 0.182 & 0.082
& 140.660 & 1.000 & 0.000 & 0.058 & 0.040\\
& UStat$^\star$
& 106.365 & 1.000 & 0.000 & 0.020 & 0.044
& 224.960 & 1.000 & 0.000 & 0.000 & 0.020\\
\addlinespace[2pt]
Covariance
& Proposed
& 3.000 & 0.995 & 0.995 & 1.848 & 0.963
& 3.000 & 0.995 & 0.995 & 2.078 & 0.979\\
& gSeg$^\star$
& 46.045 & 1.000 & 0.000 & 2.142 & 0.162
& 95.900 & 1.000 & 0.000 & 2.233 & 0.080\\
& kerSeg$^\star$
& 47.550 & 0.995 & 0.000 & 1.161 & 0.282
& 89.380 & 1.000 & 0.000 & 0.827 & 0.274\\
& UStat$^\star$
& 25.165 & 0.325 & 0.000 & 2.326 & 0.173
& 53.810 & 0.345 & 0.000 & 2.339 & 0.171\\
\addlinespace[2pt]
Marginal
& Proposed
& 2.985 & 0.860 & 0.860 & 2.491 & 0.933
& 3.000 & 0.920 & 0.920 & 2.515 & 0.969\\
distribution
& gSeg$^\star$
& 54.860 & 1.000 & 0.000 & 1.887 & 0.100
& 115.130 & 1.000 & 0.000 & 1.848 & 0.047\\
& kerSeg$^\star$
& 69.290 & 1.000 & 0.000 & 1.520 & 0.079
& 140.255 & 1.000 & 0.000 & 1.410 & 0.040\\
& UStat$^\star$
& 105.085 & 1.000 & 0.000 & 0.380 & 0.045
& 223.210 & 1.000 & 0.000 & 0.257 & 0.020\\
\addlinespace[2pt]
Joint
& Proposed
& 2.995 & 0.960 & 0.960 & 3.217 & 0.932
& 2.995 & 0.975 & 0.975 & 3.428 & 0.963\\
distribution
& gSeg$^\star$
& 27.335 & 1.000 & 0.000 & 0.972 & 0.362
& 61.310 & 1.000 & 0.000 & 0.902 & 0.167\\
& kerSeg$^\star$
& 13.580 & 0.375 & 0.000 & 4.566 & 0.341
& 21.945 & 0.380 & 0.000 & 4.712 & 0.346\\
& UStat$^\star$
& 1.535 & 0.000 & 0.000 & 7.889 & 0.144
& 1.520 & 0.000 & 0.000 & 11.094 & 0.120\\
\bottomrule
\end{tabular}
\end{table}

\clearpage

\subsection{Sensitivity analysis}
\label{Subsection: Discussion on the parameters}

We examine how the projection radius $R_0$ affects detection, and report the sensitivity of our method with respect to $m$ and $h$. All the discussions and sensitivity tests are conducted under the LS DGP.

\begin{figure}[h!]
\centering
\includegraphics[width=1.0\textwidth]{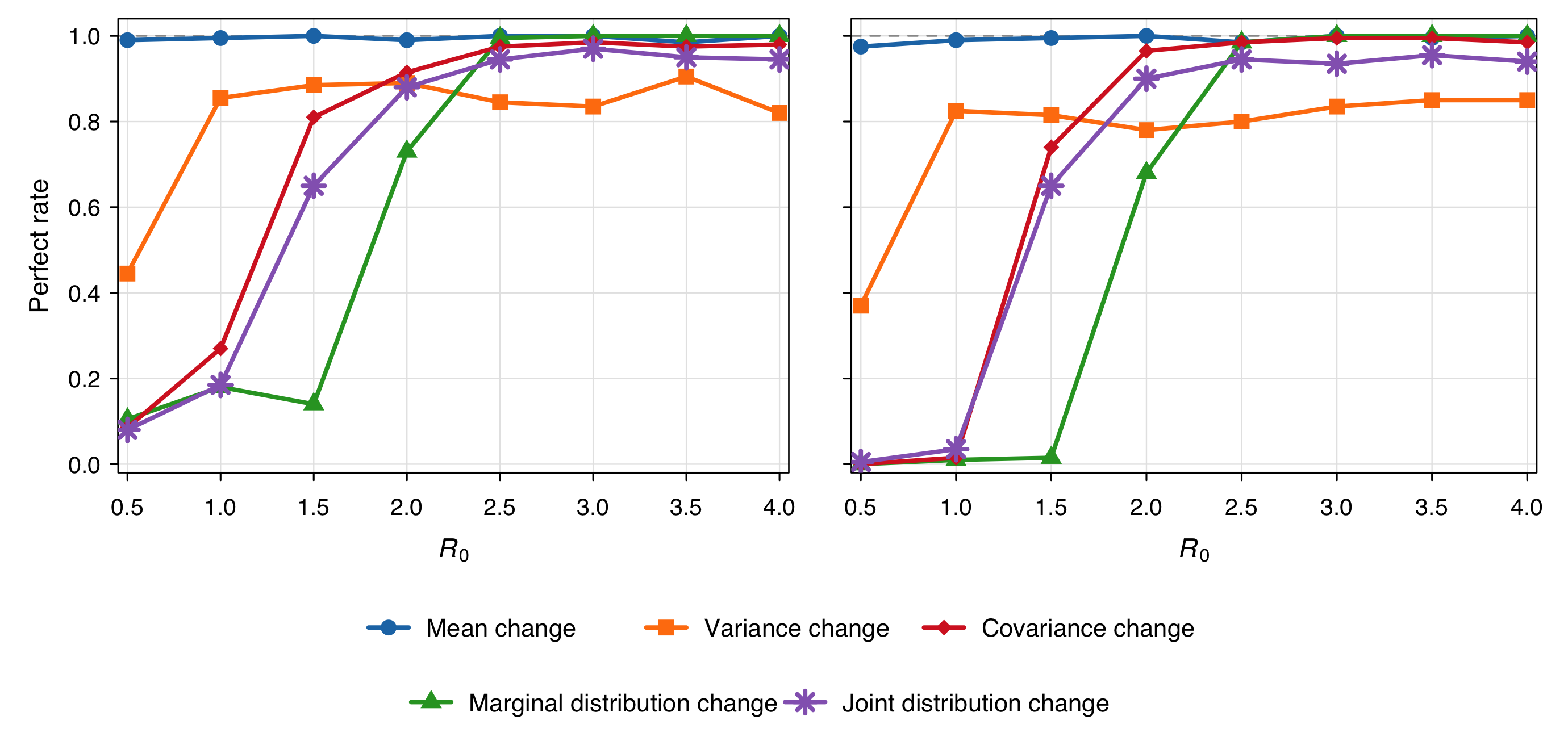}
\caption{Perfect detection probability as a function of $R_0$ for the one-change alternatives under the LS DGP with $n=400$ (left) and $n=800$ (right).}
\label{fig:ls-r0-perfect}
\end{figure}

In Algorithm~\ref{alg:randomprojection}, $R_0$ controls the frequency range over which the empirical characteristic functions are queried. The interpretation of the projection radius $R_0$ depends on the scale of the observed variables. Since rescaling a coordinate changes the effective frequency range of its characteristic function, variables are typically standardized before applying the proposed procedure so that different coordinates are comparable. Small values of $R_0$ emphasize behavior near the origin, which is often sufficient for mean and variance changes but may be less informative for covariance or shape changes. Figure~\ref{fig:ls-r0-perfect} illustrates this distinction under the LS DGP for one-change alternatives. It shows that mean and variance changes are already well captured at small or moderate regions, while covariance and distributional-shape changes benefit from a larger search radius. 

This pattern is consistent with the local behavior of characteristic functions near the origin. At a change from $P$ to $Q$, let $g(\bm t)=\varphi_Q(\bm t)-\varphi_P(\bm t)$, where $\varphi_P(\bm t)=\mathbb E_P\exp(i \bm t^\top \bm X)$.  For a sparse unit direction $\bm u$ and $\bm t=a\bm u$, a second-order expansion gives
\[
g(a\bm u)
=
ia\{\mathbb E_Q(\bm u^\top \bm X)-\mathbb E_P(\bm u^\top \bm X)\}
-\frac{a^2}{2}
\{\mathbb E_Q(\bm u^\top \bm X)^2-\mathbb E_P(\bm u^\top \bm X)^2\}
+\o(a^2),
\]
provided that the corresponding moments exist. Hence, mean and variance changes typically generate detectable signals at low frequencies. In contrast, the distributional changes considered in our simulations have identical first and second moments, causing the leading low-frequency terms to vanish. Such changes may benefit from frequencies farther from zero, where higher-order distributional differences can generate stronger signals. In addition, the pure cross-coordinate covariance changes considered here require projections involving multiple coordinates, since coordinate-wise projections do not capture cross-coordinate dependence. Therefore, after standardization, a larger $R_0$ can be beneficial for these alternatives because it allows the random projections to explore a broader frequency range.

\begin{figure}[h!]
\centering
\includegraphics[width=0.82\textwidth]{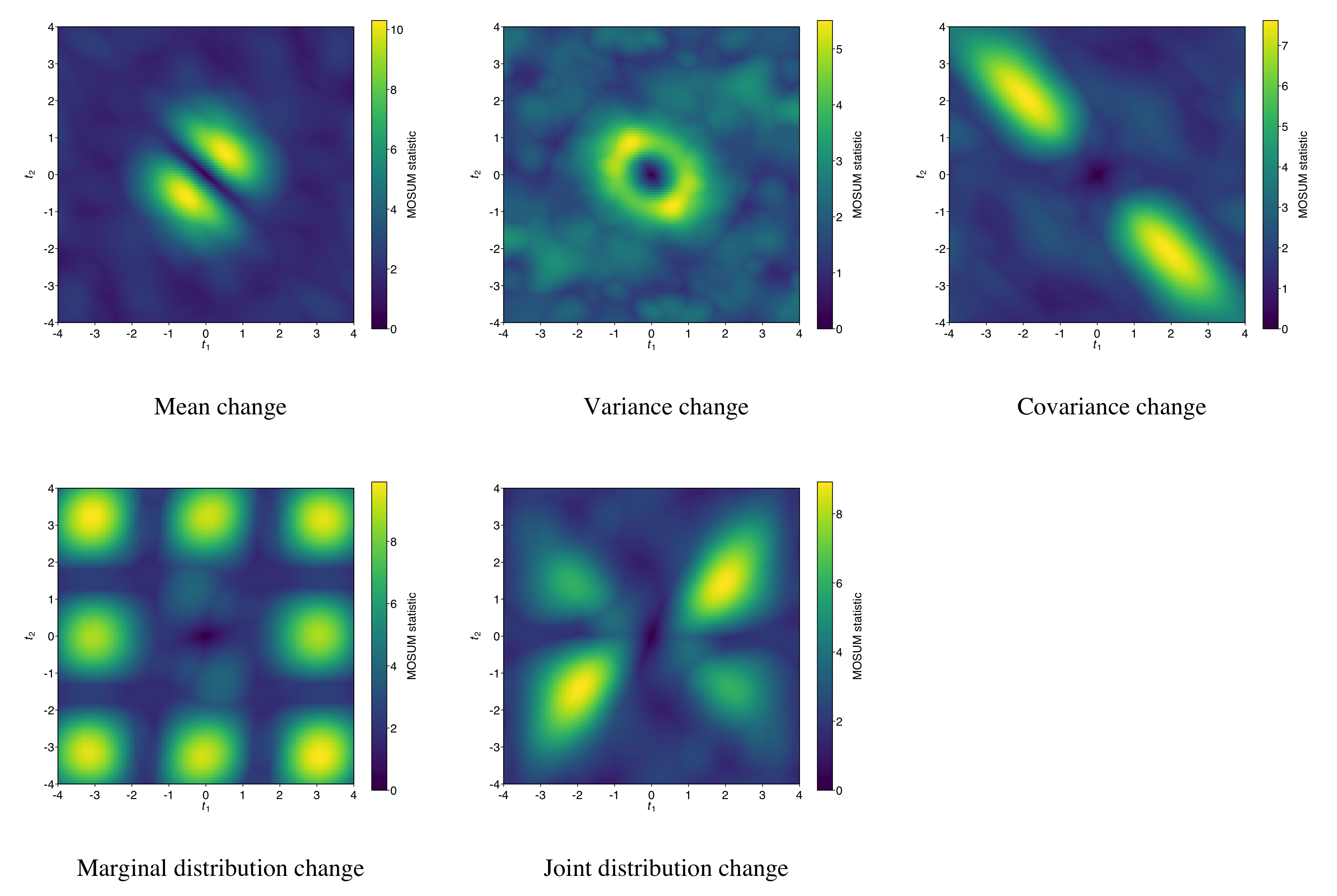}
\caption{Observed MOSUM statistic over the two-dimensional frequency plane under the one-change LS setting with $n=800$ and $p=160$. The coordinate pairs are $(1,2)$ for the first four alternatives and $(2,s_1+1)=(2,49)$ for the joint distributional change.}
\label{fig:ls-R0-sensitivity}
\end{figure}

Figure~\ref{fig:ls-R0-sensitivity} plots the observed MOSUM statistic over a two-dimensional frequency grid under the one-change LS setting. We use coordinates $(1,2)$ for the first four alternatives and $(2,s_1+1)=(2,49)$ for the joint distributional change. This figure also illustrates the differences in signal strength across alternatives with different change types, and shows why $R_0$ should not be made unnecessarily large. A larger radius enlarges the frequency region from which random projections are sampled and can waste projections in regions with weak signal. This trade-off is also reflected in Theorem~\ref{THM:computation_complexity}, which requires that $J$ increase with $R_0$. In practice, after standardizing the coordinates, we recommend using a moderate projection radius, with $R_0$ typically chosen around $3$--$4$.

\begin{figure}[h!]
\centering
\includegraphics[width=1.0\textwidth]{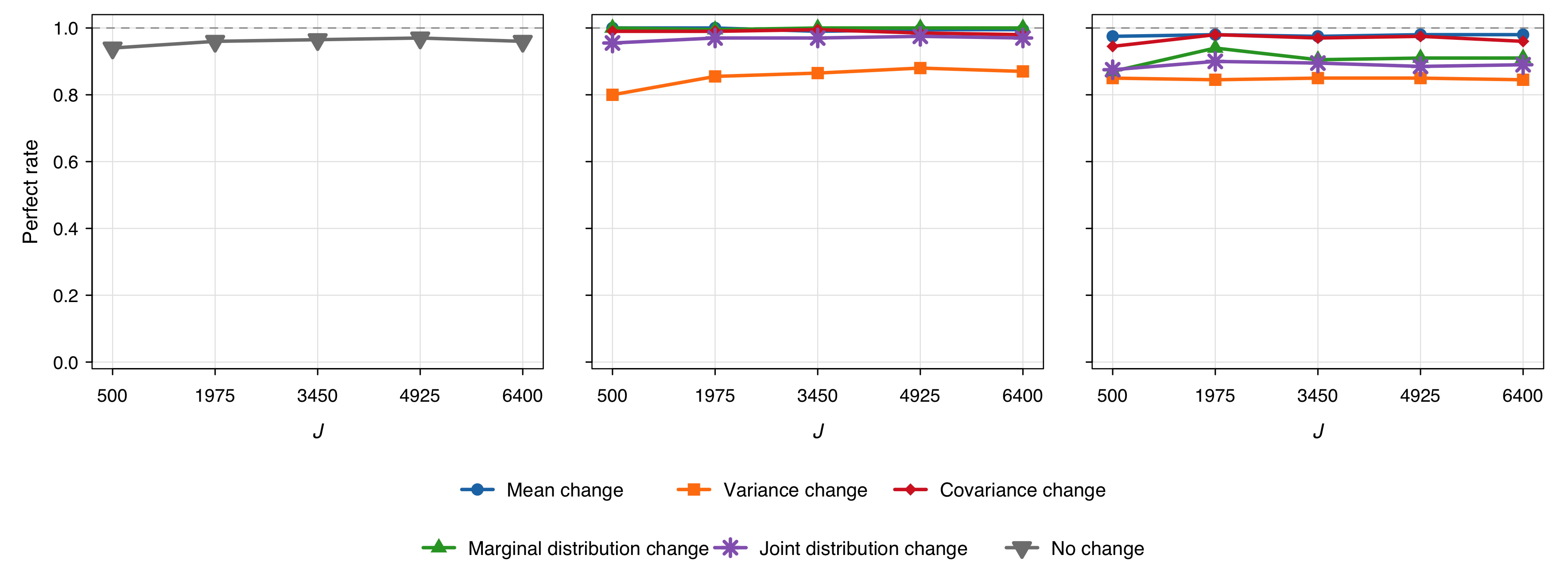}
\caption{Perfect detection probability as a function of $J$ under the LS DGP with no change (left), single change (middle), and three changes (right). $n=400$, $p=80$.}
\label{fig:ls-J-sensitivity}
\end{figure}

\begin{figure}[h!]
\centering
\includegraphics[width=1.0\textwidth]{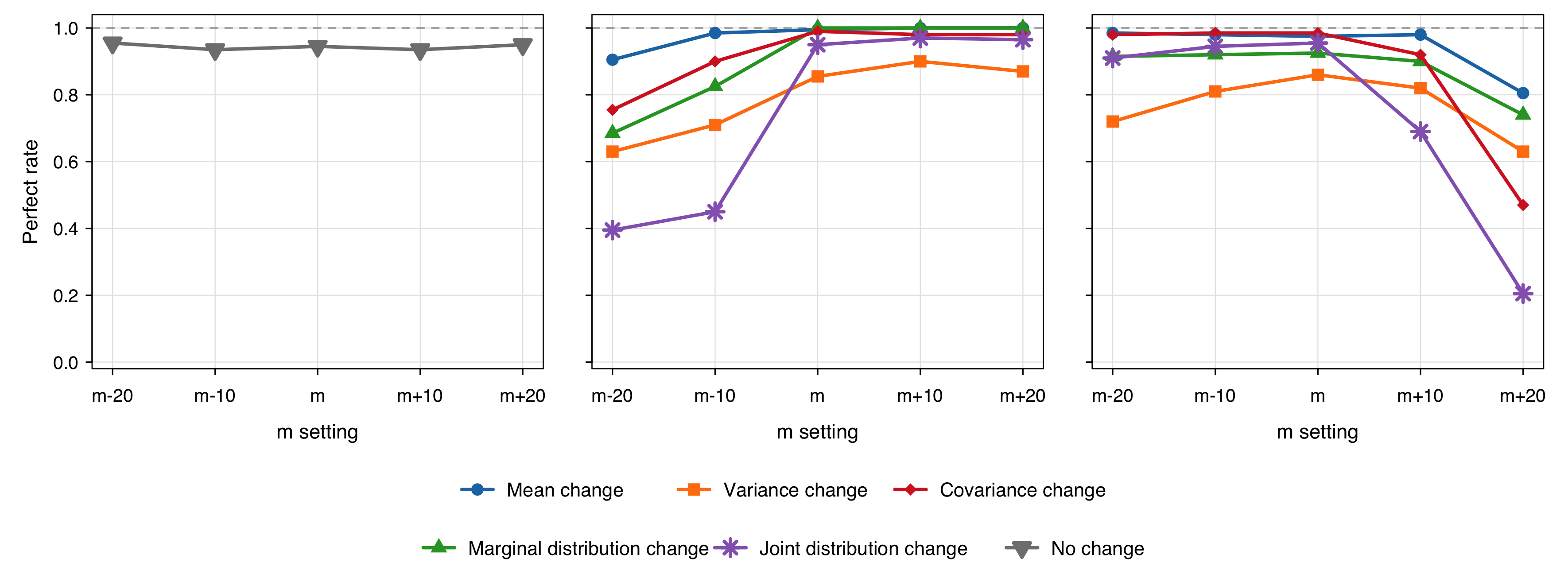}
\caption{Perfect detection probability as a function of $m$ under the LS DGP with no change (left), single change (middle), and three changes (right). $n=800$, $p=160$.}
\label{fig:ls-m-sensitivity}
\end{figure}

\begin{figure}[h!]
\centering
\includegraphics[width=1.0\textwidth]{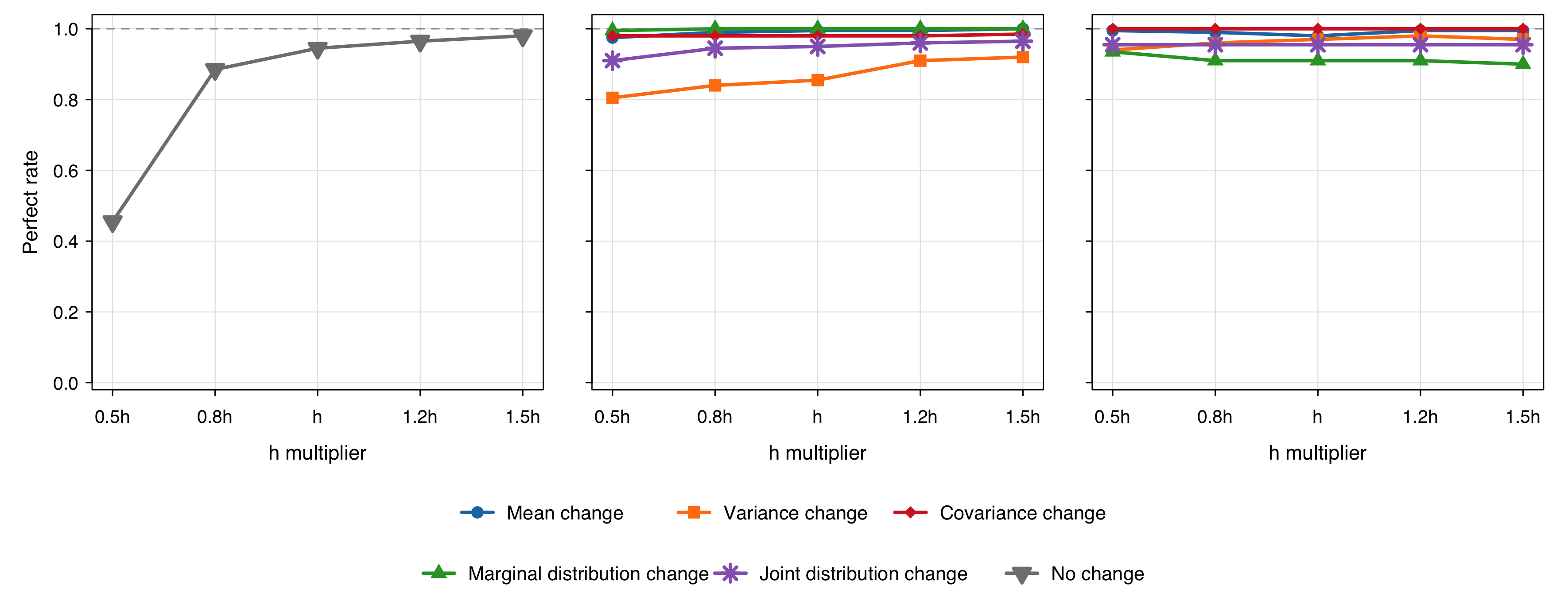}
\caption{Perfect detection probability as a function of $h$ under the LS DGP with no change (left), single change (middle), and three changes (right). $n=800$, $p=160$.}
\label{fig:ls-h-sensitivity}
\end{figure}

Finally, Figures~\ref{fig:ls-J-sensitivity}-\ref{fig:ls-h-sensitivity} assess sensitivity to $J$, $m$ and $h$ under the LS DGP. The results show that the procedure is broadly stable around the selected tuning values. Very small $h$ tends to inflate false discoveries under the null and can reduce power for distributional changes, while moderate and larger \(h\) perform well. For $m$, degradation mainly occurs when $m$ is too large for closely spaced three-change alternatives.

\clearpage

\section{Application}
\label{sec:realdata}

We apply the proposed method to the public Neuropixels recordings studied by \citet{stringer2019spontaneous}. The data contain spike times and recording locations from eight simultaneous probes across nine brain regions in awake mice, which is available at \url{https://doi.org/10.25378/janelia.7739750.v4}.  
We consider all $p=265$ recording units assigned to the hippocampal formation (HPF) in mouse Krebs and retain their common available recording interval of approximately 21.1 minutes. Each observation records the spike counts of these units in a $1/30$-second bin, giving $n=37985$ observations. Reported times are relative to the start of the published spontaneous-activity period.  

\begin{figure}[h!]
    \centering
    \includegraphics[width=\textwidth]{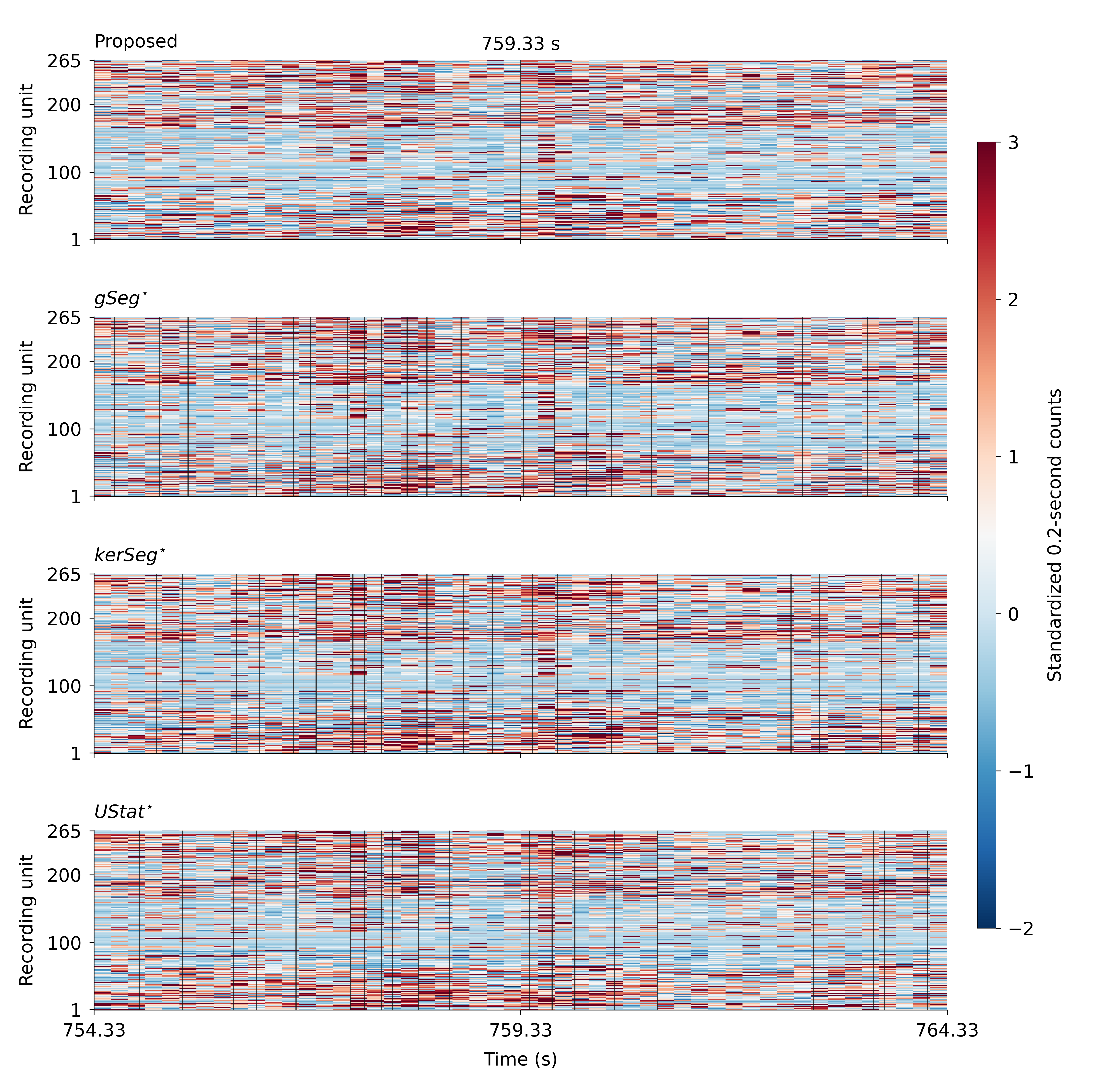}
    \caption{
    Neural activity and estimated change points around 759.33 seconds. Each panel shows the same 265 units, with black lines marking estimates from the corresponding method. Counts in 0.2-second bins are standardized within each unit over the full recording (only for visualization).
    }
    \label{fig:neural_comparison}
\end{figure}

Following the numerical simulations, we use $k_0=2$, $R_0=3$, $J=\lfloor 3.75p\rfloor+200=1193$, $\eta=0.1$, and $B=200$. We set $h=15$ and $m=300$, so each local comparison uses 10 seconds before and after a candidate location. At the nominal $5\%$ level, our method detects 50 change points. For comparison, $gSeg^{\star}$, $kerSeg^{\star}$, and $UStat^{\star}$ detect 1447, 1590, and 890 change points, respectively. In Figure~\ref{fig:neural_comparison}, the competing methods report several closely spaced change points within the same interval where our method detects one. The denser detections by the  competing methods are reminiscent of the over-detection observed in our simulations under  temporal dependence. Our results are also stable under moderate changes in the hyperparameters. We vary $m\in\{200,300,400\}$ and $h\in\{10,15,20\}$ with the random projections and bootstrap multipliers held fixed. Using the 50 estimates at $(m,h)=(300,15)$ as a benchmark and the one-to-one matching within $3\log n$ time points as in the simulations, 22 detected change points have a match across all configurations.

\begin{figure}[h!]
    \centering
    \includegraphics[width=\textwidth]{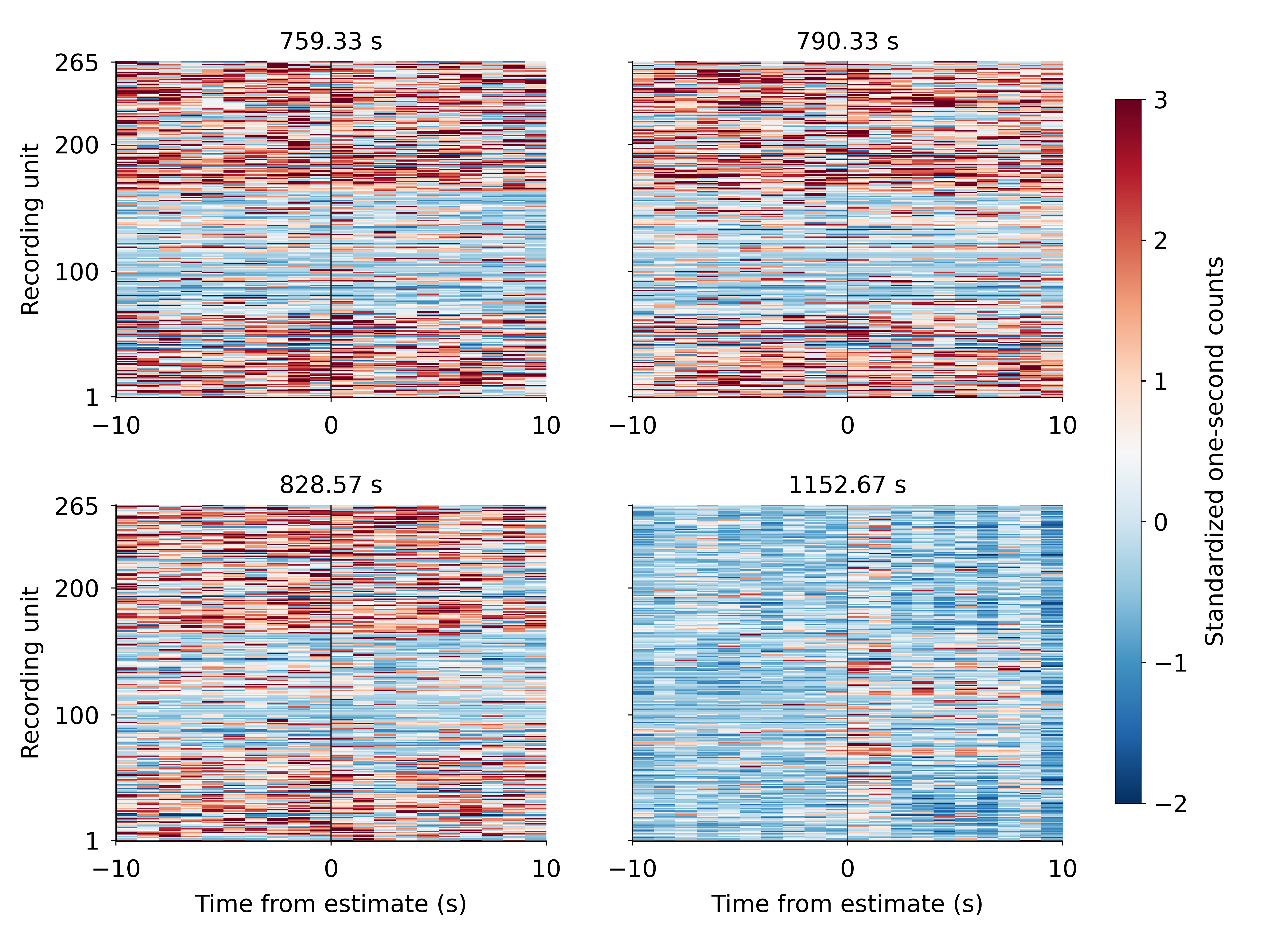}
    \caption{Activity of 265 recording units within 10 seconds of four estimated change points. The vertical line marks the estimated change point. Titles give the estimated times in seconds. Counts in 1.0-second bins are standardized using each unit's full-record mean and standard deviation (for visualization).
    }
    \label{fig:neural_overview}
\end{figure}

\begin{figure}[h!]
    \centering
    \includegraphics[width=\textwidth]{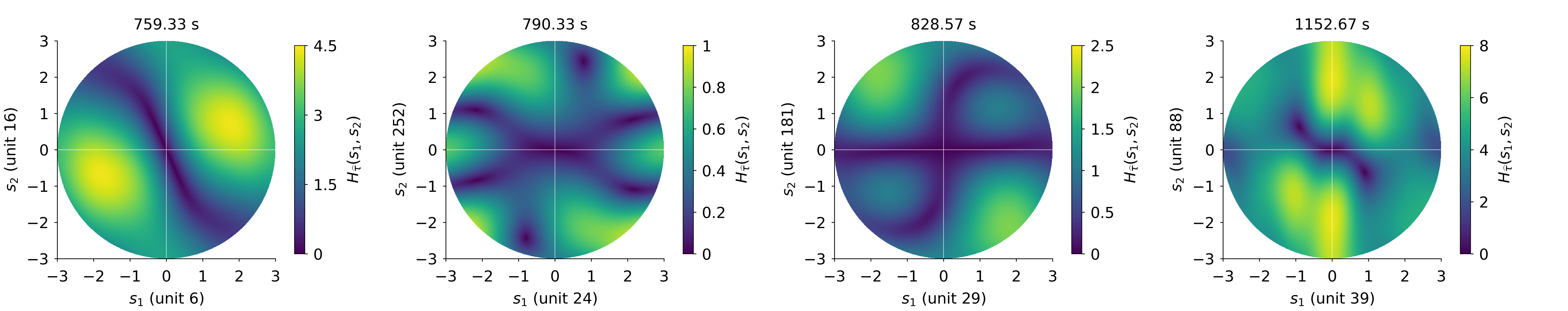}
    \caption{
    Local characteristic-function contrasts at four estimated change points. 
    }
    \label{fig:neural_cf}
\end{figure}

Figure~\ref{fig:neural_overview} displays the activity of all 265 units around four representative change points detected by our method. 
Guided by the sparse projections, we further examine selected pairs of units to see how these changes are expressed in their marginal and joint activity. For each estimated change point $\widehat{\tau}$ and some coordinates $a$, $b$, consider the local characteristic-function contrast
\[
H_{\widehat{\tau}}(s_1,s_2)
=
\left|
\sum_{i=\widehat{\tau}-m+1}^{\widehat{\tau}}
e^{\mathrm{i}(s_1X_{ia}+s_2X_{ib})}
-
\sum_{i=\widehat{\tau}+1}^{\widehat{\tau}+m}
e^{\mathrm{i}(s_1X_{ia}+s_2X_{ib})}
\right|/{\sqrt{2m}},\quad s_1^2+s_2^2\leq R_0^2.
\]
Larger contrasts on an axis indicate a stronger marginal signal in the corresponding unit, while off-axis values compare their empirical joint characteristic functions and can reflect changes in marginal distributions, dependence, or both. Figure~\ref{fig:neural_cf} shows these contrasts for the selected coordinates at the four representative change points. Together with local count summaries, they help characterize the distinct forms of change below.
\begin{itemize}
    \item {(Change in mean)} At 759.33 seconds, both axis contrasts indicate marginal changes in units 6 and 16, as in many other pairs. We compare the mean firing rates before and after this change point and find that increases in some units largely offset decreases in others, leaving the population average nearly unchanged at 14.74 and 14.75 spikes per second per unit, respectively. 
    \item {(Change in joint upper-tail behavior)} At 790.33 seconds, units 24 and 252 exhibit more frequent joint upper-tail exceedances. Using the empirical 85th percentiles pooled across the 600 bins in the two adjacent windows as thresholds, the proportion of bins where both counts strictly exceed their respective thresholds rises from 0.67\% to 3.00\%. Meanwhile, both units' sample means and variances change by less than 6\%, and their sample covariance remains nearly unchanged at 0.0152 and 0.0157.
    \item {(Change in covariance)} At 828.57 seconds, strong contrasts away from the axes and weak axis contrasts suggest a joint distributional change in units 29 and 181. We find that their empirical count covariance changes from $-0.06$ to $0.08$, while each unit's mean and variance change by less than $5\%$. 
    \item {(Change in variance)} At 1152.67 seconds, several units are active less often but produce more spikes per active bin. For example, in unit 39, the active-bin proportion falls from 51.0\% to 35.3\% while the mean count per active bin rises from 1.65 to 2.52, raising its sample variance from 0.96 to 2.25.
\end{itemize}
These examples demonstrate that changes in neural activity take diverse forms. Without prespecifying the form of change, the proposed method captures these heterogeneous changes within a unified framework. The selected sparse projections highlight potentially informative units, while the associated characteristic-function contrasts help illustrate the underlying distributional changes.

\section{Concluding Remarks}

In this paper, we developed a general framework for high-dimensional
distributional change point analysis under nonstationary temporal dependence.
The proposed method combines sparse random projections with characteristic
functions and a Gaussian multiplier bootstrap approach, allowing
general distributional changes to be detected without imposing
stationarity or piecewise stationarity on the underlying temporal dynamics.
Our theory quantifies the computational complexity of identifying informative
projection directions, establishes bootstrap validity under a general
nonstationary physical dependence framework, and provides asymptotic
false-detection control together with simultaneous guarantees for estimating
the number and locations of multiple change points. A key technical ingredient
is a Gaussian approximation for high-dimensional sparse time
series, which allows heterogeneous effective sample sizes across coordinates
and does not require the coordinatewise variances to be uniformly bounded away
from zero. Simulation studies further demonstrate the effectiveness of the
proposed procedure across a broad range of distributional changes and temporal
dependence structures.

Several directions merit further investigation. One interesting direction is
to develop statistical inference procedures for recovering the informative
support underlying the detected distributional changes. Such procedures would
provide a more refined understanding of the coordinates that drive the
distributional shifts. Another possible extension is to combine the proposed
test with alternative recursive procedures, such as binary segmentation,
which we leave for future research.

\section*{Acknowledgments}
The authors would also like to thank Mr.\ Yang-Guan-Jian Guo for his contribution in developing an early version of the Gaussian approximation results  in this paper. 

\bibliographystyle{apalike}
\bibliography{ref}

\end{document}